\documentclass[letterpaper]{article} % DO NOT CHANGE THIS
\usepackage{aaai2026}  % DO NOT CHANGE THIS
\usepackage{times}  % DO NOT CHANGE THIS
\usepackage{helvet}  % DO NOT CHANGE THIS
\usepackage{courier}  % DO NOT CHANGE THIS
\usepackage[hyphens]{url}  % DO NOT CHANGE THIS
\usepackage{graphicx} % DO NOT CHANGE THIS
\usepackage{amsmath, amssymb}
\usepackage{amsthm}
\usepackage{bm}
\usepackage{natbib}  % DO NOT CHANGE THIS AND DO NOT ADD ANY OPTIONS TO IT
\usepackage{caption} % DO NOT CHANGE THIS AND DO NOT ADD ANY OPTIONS TO IT
\usepackage{algorithm}
\usepackage{algorithmic}

\usepackage{newfloat}
\usepackage{listings}
\DeclareCaptionStyle{ruled}{labelfont=normalfont,labelsep=colon,strut=off} % DO NOT CHANGE THIS
\floatstyle{ruled}
\newfloat{listing}{tb}{lst}{}
\floatname{listing}{Listing}
\title{Think How Your Teammates Think: Active Inference Can Benefit \\ Decentralized Execution}
\author{
    Hao Wu\textsuperscript{\rm 1, 2}\equalcontrib,
    Shoucheng Song\textsuperscript{\rm 1, 2}\equalcontrib,
    Chang Yao\textsuperscript{\rm 1, 2},
    Sheng Han\textsuperscript{\rm 1, 2},\\
    Huaiyu Wan\textsuperscript{\rm 1, 2},
    Youfang Lin\textsuperscript{\rm 1, 2},
    Kai Lv\textsuperscript{\rm 1, 2}\thanks{Corresponding Author: Kai Lv (lvkai@bjtu.edu.cn).}
}
\affiliations{
    \textsuperscript{\rm 1}School of Computer Science \& Technology, Beijing Jiaotong University, Beijing, China\\
    \textsuperscript{\rm 2}Beijing Key Laboratory of Traffic Data Mining and Embodied Intelligence, Beijing, China
    \{wuhao\_, insis\_songsc, yaochang, shhan, hywan, yflin, lvkai\}@bjtu.edu.cn
}

\usepackage{bibentry}
\begin{document}

\maketitle

\begin{abstract}
In multi-agent systems, explicit cognition of teammates' decision logic serves as a critical factor in facilitating coordination.
Communication (i.e., ``\textit{Tell}'') can assist in the cognitive development process by information dissemination, 
yet it is inevitably subject to real-world constraints such as noise, latency, and attacks.
Therefore, building the understanding of teammates' decisions without communication remains challenging.
To address this, we propose a novel non-communication MARL framework that realizes the construction of cognition through local observation-based modeling (i.e., \textit{``Think''}). Our framework enables agents to model teammates' \textbf{active inference} process. 
At first, the proposed method produces three teammate portraits: perception-belief-action. Specifically, we model the teammate's decision process as follows: 1) Perception: observing environments; 2) Belief: forming beliefs; 3) Action: making decisions. 
Then, we selectively integrate the belief portrait into the decision process based on the accuracy and relevance of the perception portrait.
This enables the selection of cooperative teammates and facilitates effective collaboration.
Extensive experiments on the SMAC, SMACv2, MPE, and GRF benchmarks demonstrate the superior performance of our method.
\end{abstract}


\section{Introduction}

Multi-agent reinforcement learning (MARL) has garnered significant attention due to its wide applications in fields such as autonomous driving \cite{kiran2021deep}, smart grids \cite{roesch2020smart}, and transportation \cite{lee2019reinforcement}. 
% In decentralized systems, the inability of agents to acquire timely teammate information can lead to uncooperative behaviors, ultimately resulting in suboptimal policies.
In decentralized systems, the lack of cognition regarding teammates' decision logic may induce miscoordination among agents and result in suboptimal policies.
% Many MARL methods employ the centralized training and decentralized execution (CTDE) \cite{bernstein2002complexity} paradigm to improve agent collaboration \cite{lowe2017multi,rashid2020monotonic,yu2022surprising,yao2025general}. However, during the execution of CTDE, agents can only access partial observations. This limitation introduces uncertainty about teammates, which can trigger uncooperative actions and result in inefficient policies.
% Moreover, incomplete information about teammates prevents agents from accurately evaluating the impact of their actions on coordination.
% 在分布式系统中，如果智能体无法及时获取队友信息，这会导致智能体之间无法进行很好的协作，从而陷入局部最优。

\begin{figure}[t]
\centering
\includegraphics[width=1\columnwidth]{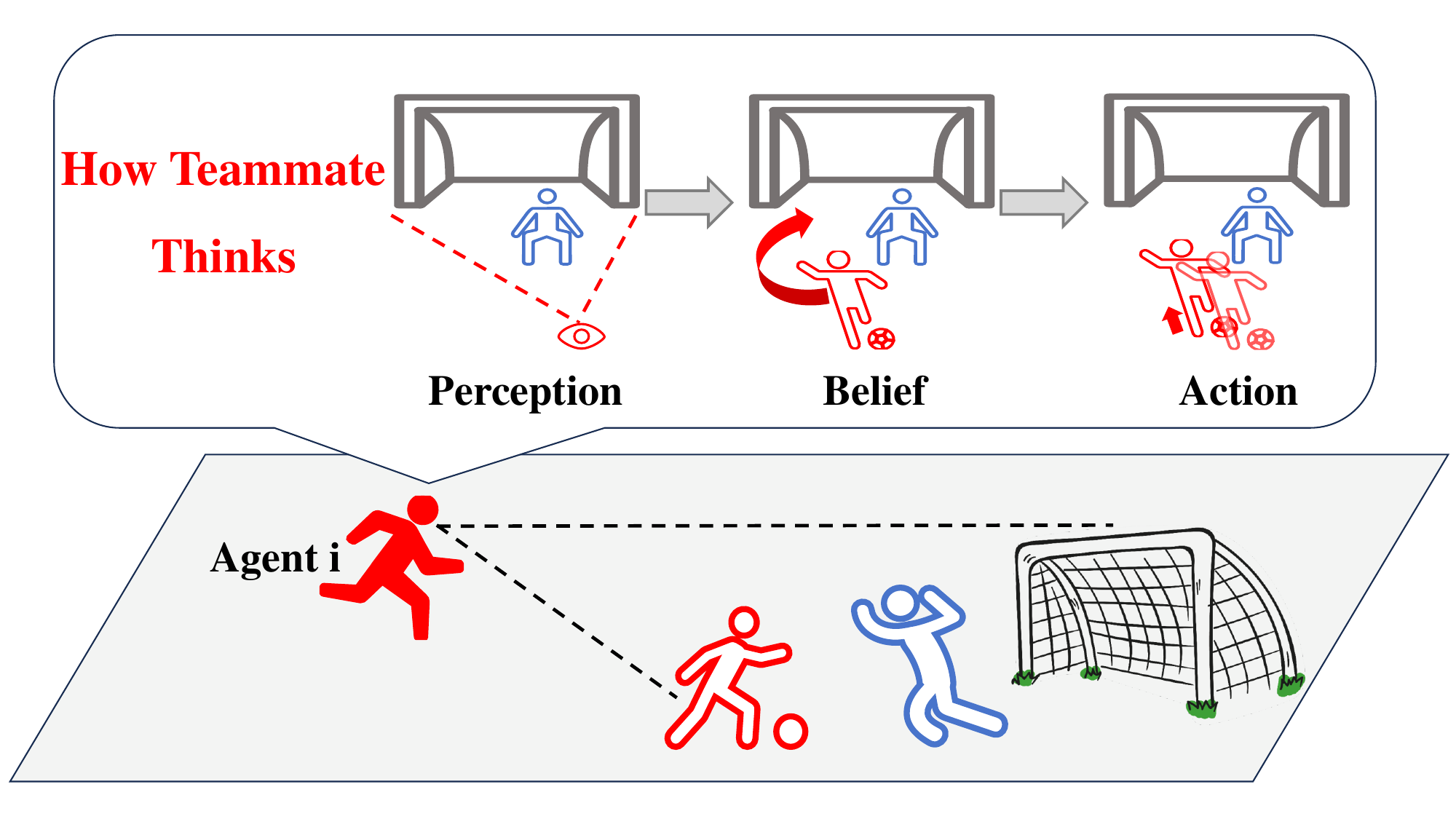}
\caption{Modeling the Active Inference Process of the Teammate. In this scenario, agent \(i\) models the perception-belief-action involved in its teammate's active inference process when facing the goalkeeper. This allows \(i\) to obtain its teammate's decision-relevant information and achieve effective collaboration.}
\label{fig: intro}
\end{figure}
% Tell your teammates how they act

To address the mentioned issue, one intuitive approach is ``\textit{\underline{Tell} agent how its teammates act}", which can be implemented through communication mechanisms. The communication (i.e., ``\textit{Tell}") methods can facilitate the understanding of teammates' behaviors by exchanging decision-relevant messages \cite{wang2019learning,yuan2022multi,sun2023cifar,sun2024t2mac}.
However, communication effectiveness may be limited under some conditions, including limited bandwidth, high latency, and significant noise \cite{9466501,9597491,Song_Lin_Han_Yao_Wu_Wang_Lv_2025}. 
% Therefore, we consider obtaining teammate information without communication. 
Therefore, we explore the method for developing the cognition of teammates' decisions in the absence of communication.

% 因此，我们考虑在无通信情况下获取队友信息。

% 相比于被动被告知队友的消息，我们认为智能体应该主动Think how your teammates think来获取消息。然而，现有的智能体建模方法只考虑对队友的决策进行部分建模且无法适用于CTDE。
% \textcolor{red}{Rather than ``\textit{\underline{Tell} agent how its teammates act}", a reasonable solution is ``\textit{\underline{Think} how your teammates think}".} 
In contrast to directly telling an agent how its teammates act, we advocate that agents could engage in ``\textit{\underline{Think} how your teammates think}".
% \textcolor{red}{
Specifically, the agent actively builds comprehension of teammates' decisions.
% 受主动推理的启发，智能体的决策过程可以看作是主动推理过程。这个过程包括感知、信念和动作三部分。智能体首先感知当前环境，然后根据感知形成自己的认知信念，之后结合感知和信念进行决策。我们认为受控智能体应该主动建模队友的主动推理过程来
% understand how they think.
% 具体来说，智能体应该主动获取有关队友决策的相关信息,而不是被动告知。
% \textcolor{red}{
% Inspired by active inference \cite{friston2016active}, we regard the teammate's decision-making as the active inference process. 
% Within this process, the teammate continuously updates cognitive beliefs based on the perception of the environment. By combining current perceptions with beliefs, the teammate makes optimal decisions. }
% 尽管一些方法仅使用局部观测对智能体进行建模，但他们仅训练一个受控智能体，同时保持队友的策略是预训练好的，同时对对手建模而不考虑队友。
% 为了获得队友主动推理过程中的感知、信念和动作，一种直接的方法是智能体建模。图一提供了一个例子来说明这一思考过程.然而，现有的建模方法无法实现这一点。一方面，一些方法只部分建模其他智能体决策过程，例如意图或策略。由于模型和真实情况的差异，这种部分建模会导致不准确性。
% 智能体建模的方法可以通过构建队友的信念、动作或意图来了解队友的部分思考过程。
% \textcolor{red}{To achieve the above idea,} one approach is agent modeling. 
% To achieve this, agent modeling serves as a direct approach.
To achieve this, the direct method is to model teammates' decision process.
% agent modeling methods can reconstruct teammates' decision-making logic through modeling their beliefs, actions, or intentions.
However, existing agent modeling methods fail to fulfill this.
% Existing agent modeling methods focus on partial aspects of teammates, such as policies or intentions. However, such unilateral modeling inevitably leads to inaccuracies due to discrepancies between the model and the actual situation. Meanwhile, existing methods are incompatible with CTDE.
On the one hand, some methods rely on access to other agents' trajectories during the modeling  \cite{rabinowitz2018machine,zintgraf2021deep}, which is unavailable during decentralized execution.
% On the one hand, some methods only model incomplete decision processes of other agents, such as intentions or policies \cite{he2016opponent,rabinowitz2018machine,zintgraf2021deep}. Nevertheless, due to discrepancies between the model and the actual situation, relying on incomplete modeling may cause inaccuracies.
On the other hand, 
% some methods rely on direct access to other agents' trajectories or behaviors during the modeling phase \cite{rabinowitz2018machine,zintgraf2021deep}, which is unavailable during execution. Although some methods model agents using only local observations, they 
some methods typically only enable a single agent to model other agents that possess fixed parameters 
% they control a single independent agent, where teammate policies are typically pre-trained.
% they treat controlled agents as independent learners with pre-trained teammate policies 
% 因此，需要构建一个基于CTDE且仅使用局部观测来实现对队友主动推理过程建模的框架。同时，只建模智能体决策的一部分容易产生建模偏差。
\cite{xie2021learning,papoudakis2021agent,yu2024opponent}. This configuration imposes an upper bound on the system's collaborative efficiency, preventing the team from learning more optimal policies.
% Conversely, CTDE requires joint training of all cooperative agents for complex task completion.
% This contradicts the CTDE paradigm, which requires training all teammates.
% 这会导致整个系统的协作效能存在上限，无法使团队学习到更好的策略。
% Conversely, CTDE requires training multiple controlled agents, including teammates. 
% Meanwhile, partial modeling of agents' decision processes may introduce biases.
% On the other hand, the aforementioned methods focus on partial aspects of teammates, such as policies or intentions. Nevertheless, such unilateral modeling may lead to inaccuracies due to discrepancies between the model and the actual situation.
% Moreover, these methods only model incomplete decision processes of other agents, which may cause inaccuracies due to discrepancies between the model and the actual situation.
% Therefore, these methods fail to be extended to CTDE and then mitigate the uncertainty.
% \textcolor{red}{Moreover, these methods only model incomplete decision processes of other agents, potentially causing inaccuracies from model-reality mismatches.}
Furthermore, these methods only model incomplete decision components (e.g., behaviors or intentions), risking inaccuracies from discrepancies between the model and the actual situation.
% To address this, we seek to model teammates' complete decision-making process to reduce modeling inaccuracies solely based on local observations in CTDE.
To address this, we rely solely on local observations to model teammates' complete decision processes, reducing modeling inaccuracies during decentralized execution.
% To address this, we seek to develop a local observation-based framework for modeling teammates' complete decision processes, thereby reducing inaccuracies.
% }

% \textcolor{red}{
% In order to achieve \textit{``Think''}, we need to model teammates' complete decision-making process.
Inspired by human brain decision-making mechanisms and active inference theory \cite{friston2016active}, we model the teammate's decision process as an active inference process comprising perception-belief-action. In this process, the teammate perceives environments, forms beliefs, and then takes actions by integrating perceptions and beliefs.
% We aim to construct the teammates' comprehensive active inference process to understand how they think. 
% 因此，我们想要通过主动建模的方法来对了解队友的感知、信念和动作。
% We argue that controlled agents should actively model teammates' comprehensive active inference process to understand how they think.
% Thus, active thinking requires agents to acquire and maintain a fundamental understanding of teammates' perception-belie-action.
% Consequently, we want to acquire and maintain a fundamental understanding of teammates' perception-belief-action through active modeling (i.e., \textit{``Think''}).
Consequently, we employ local observation-based modeling (i.e., \textit{``Think''}) to acquire teammates' perception-belief-action (i.e., \textit{``How your teammates think''}).
% Figure \ref{fig: intro} provides an example of how an agent thinks about how its teammate thinks. 
% Figure \ref{fig: intro} provides an example to demonstrate this thought process.
Figure \ref{fig: intro} provides an illustrative example of this modeling process.
% \textcolor{red}{requirement. }
% In contrast, the CTDE paradigm requires interactive training and coordination for complex tasks. 
% Therefore, the aforementioned modeling methods fail to effectively address the uncertainty issue during the execution.
% This independent modeling fails to capture inter-agent collaboration patterns, consequently degrading cooperative performance.
% 为了解决该问题，我们尝试将建模方法引入执行阶段，建模其他智能体的信息以缓解在执行阶段的非平稳性问题。之前的智能体建模方法[He et al., 2016]，Rabinowitz et al., 2018]，OP[Ma et al., 2021b]，IJCAI直接使用队友的轨迹信息或行为信息来建模队友的观测、策略、意图。然而，这些建模方法都是在能够直接获取建模智能体完整信息的强假设下实现的。但对于智能体来说，使用完整的其他智能体的历史信息在CTDE的测试过程中也是无法实现的。因此以上提到的建模方法无法扩展到CTDE框架中。同时，由于建模信息与真实情况之间存在差异，单方面建模且使用准确性较低的建模结果可能会导致智能体产生错误的认知，影响协同性能。
% 受到主动推理的启发，我们考虑完全合作多智能体任务，想要通过构建队友完整的主动推理过程来理解队友是怎么想的，以此来缓解执行阶段的决策不确定性。我们基于CTDE仅使用局部观测并对每一个友方智能体进行训练。智能体的决策过程可以看作是一个主动推理过程。在这一过程中，智能体通过观测感知状态信息不断更新对环境的认知信念，并根据当前的感知和信念，做出最佳的决策。根据这一思想，我们提出一个全新的MARL建模队友主动推理过程的框架。该框架包含两个部分1）我们在执行过程中仅使用智能体的局部信息全面建模队友的感知-信念-动作，从而弥补执行阶段队友信息的缺失 2）对于建模的准确性问题，我们提出双重过滤机制，通过剔除准确性低、相关性差的画像，增强建模画像的可靠性和决策的精确性。

% 由于对手动作在训练时无法获取,因此我们只考虑对队友进行建模。在这篇文章中，我们考虑完全协作的多智能体任务，propose a novel framework for modeling the \textbf{A}ctive \textbf{I}nference of teammates in \textbf{M}ARL (AIM)

% In this paper, we consider fully cooperative multi-agent tasks and follow the CTDE, training each allied agent using only local observations. Specifically, 
In this paper, we propose a novel non-communication framework for modeling the \textbf{A}ctive \textbf{I}nference of teammates in \textbf{M}ARL (AIM).
The framework consists of two parts: At first, we develop a modeling method to model teammates' three portraits: perception-belief-action, solely based on local observations.
% Specifically, we initially model teammates' perceptions to understand the world. 
% Subsequently, we model the specific beliefs of teammates to understand their cognitive processes. Finally, we model teammates' actions and update the perception and belief portraits by minimizing the discrepancy between predicted and actual actions. 
Meanwhile, the perception and belief portraits are optimized by minimizing the discrepancy between predicted and actual actions. 
Then, we propose a dual-filter mechanism to enhance teammates' cognition utilization.
This mechanism features selective collaboration by choosing teammates whose modeled portraits have high accuracy. Additionally, by considering the perception relevance among agents, we adopt an attention module to dynamically integrate teammates' belief portraits, thereby optimizing the decision process. 
% \textcolor{red}{Since the actions of opponents are inaccessible during the training phase, we focus solely on modeling teammates. }
% 由于建模画像和真实情况之间存在差异以及使用所有的建模画像是不现实的，因此我们在提出的框架中考虑选择性合作，选择建模画像准确性高的队友作为潜在合作伙伴，并考虑智能体间感知历史相关性使用注意力模型动态融合队友信念画像来优化决策。
% 在这篇文章中，我们将建模视角从主体智能体转移到队友，构建队友的主动推理框架。我们首先从根据视角转换建模队友观测的子集；然后结合历史信息和队友特定ID，建模队友的信念；最后，基于前两部分的建模结果，建模队友的动作。
% % 另外，由于观测建模结果与真实值之间存在差异，我们根据观测建模的视角转换相关性质提出评价观测建模结果准确性的方法，并选取使用top_k个准确率最高的队友索引作为合作伙伴。
% 由于在自身决策过程中使用全部队友建模信息是不现实的，相比于短期的观测和动作，我们使用能够了解队友未来规划的信念来辅助决策，并使用attention模型融合队友信念。我们提出的方法在smac,smacv2和grf中效果明显，甚至在一些较难的任务中超过了通信算法。
Our proposed method demonstrates significant improvement in tasks within SMAC \cite{samvelyan2019starcraft}, SMACv2 \cite{ellis2024smacv2}, MPE \cite{mordatch2018emergence} and GRF \cite{kurach2020google}.

% 另外，观测建模本身视角转换的结果，队友智能体的建模结果是在队友的视角下得到的，本身是队友观测的局部，与其他智能体无关。同时由于观测建模结果与真实值之间存在差异，我们根据观测建模的不相关性，提出满足自己是保真的、评价是相似的、高相似性赋予高分三种性质的评价观测建模结果准确性的方法，并选取使用top_k个准确率最高的建模结果。由于建模的是队友的整个主动推理模型，在自身决策过程中使用全部的队友建模信息是不现实的。相比于短期的动作，我们使用能够了解到队友未来规划的信念来辅助决策，使用attention模型融合队友信念。我们提出的方法在smac,smacv2和grf中效果明显，甚至在一些较难的任务中超过了通信算法。

% 我们的贡献概述如下：
Our contributions are outlined as follows:
% 1、我们从"信息告知"(Tell)转变为"主动建模"(Think)，使智能体能够通过主动建模自主获取队友信息。
% 1、我们提出队友主动推理框架来主动建模队友感知、信念和动作，使智能体了解更多队友信息，缓解MARL中的非平稳性。
% 2、我们根据感知画像评价建模准确性，并融合与自身相关的信念建模画像来了解队友的长期规划。
% 3、我们在SMAC,SMACv2,GRF上进行了实验。PA的效果要优于baseline算法。
% \textit{\underline{Tell} agent how its teammates act}", a reasonable solution is ``\textit{\underline{Think} how your teammates think}"
\begin{itemize}
    \item We replace ``communication (i.e., \textit{Tell})" with ``modeling (i.e., \textit{Think})", enabling agents to construct the cognition of teammates' decision logic without communication during decentralized execution.
    \item We propose an active inference framework to model teammates' three portraits: perception-belief-action, to understand how they think.
    % This approach enables agents to understand how teammates think and alleviate uncertainty in execution.
    \item We introduce a dual filter that leverages the accuracy and relevance of perception portraits to select cooperative teammates.
    \item We conduct experiments on SMAC, SMACv2, MPE, and GRF. The results show that our method achieves optimal or near-optimal performance in most scenarios.
\end{itemize}

\section{Related Works}
% 解决分散执行过程中智能体间协作问题：共识方法，智能体建模方法，角色方法
% 多智能体间合作
% 由于智能体在CTDE执行阶段单独决策，环境仍然存在非平稳性的问题，为了解决该问题，之前的MARL研究者提出了一系列工作，例如COLA，IROS，3等等通过在执行阶段引入共识作为全局指导来辅助决策，其中MAGI通过推断未来状态中具有高价值的状态作为智能体的共识目标状态并将其作为内在奖励促进协作，RA-CTDE通过周围智能体推断中心智能体动作倾向并设计动作内在奖励集成到CTDE框架。但是，环境的变化使智能体需要不断调整共识，这一过程计算开销大且计算复杂。
% 另外，一些通信算法，如NDQ,TMC,TarMAC,1,2,3,4,5从Tell your teammates how they act角度出发，设计了通信网络使智能体能在执行阶段进行信息交互，T2MAC通过证据理论来实现智能体间的稀疏通信。但在一些现实场景中，高噪声、高延迟、低带宽的限制使通信算法无法达到很好的效果。我们从Think How Your Teammates Think方向思考，建模队友的主动推理模型，并考虑合作队友筛选，不需要额外的通信网络，避免了这些环境限制。

% 基于角色的方法ROMA，RODE，SIRD，COPA等等通过将智能体分成不同的角色来完成协作任务，但这些方法通常存在将角色固定化、忽视团队动态变化和弱化CTDE约束的问题，导致无法有效应对智能体行为和任务需求的复杂性。我们的方法不需要设计内在奖励，同时是智能体的策略学习是在强CTDE约束下进行的。
% 智能体建模
% 基于智能体建模的方法往往通过对智能体的观测，行为或意图进行建模来补充部分可观测导致的信息缺失，例如1,2,3，4，8基于心智理论和贝叶斯推理对对手的心理状态和信念进行建模，5，6通过使用对手的观测和观察到的动作信息对对手的策略进行推断，7，8将对手的策略建模为潜在分布，9则通过离线和在线的结合应对对手策略的变化。然而这些方法都基于能够直接获取队友的相关信息的强假设。这在执行阶段是无法实现的。一些方法虽然仅通过局部观测下对对手进行建模，但他们将每个智能体看作是一个独立的个体，队友策略往往是预训练好的。然而在CTDE范式中，。在友方智能体数量较多时，该方法面临内存挑战。相比之下，我们基于CTDE，只根据局部信息从感知，信念，动作三个方面建模队友的主动推理模型,来了解队友的行为逻辑，从而缓解在执行阶段由于队友信息缺失导致的智能体决策不确定性。
\subsection{Communication in MARL}
% \noindent\textbf{Collaboration in CTDE Execution}
% Since the agent makes decisions independently during the CTDE execution phase, agents still face the problem of non-stationarity. To address this issue, previous research in MARL has proposed a series of methods, such as \cite{xu2023consensus,ruan2023learning,feng2024hierarchical,wang2024reaching,zhang2024intrinsic} which assist decision-making by introducing consensus as global guidance during execution. MAGI \cite{wang2024reaching} infers high-value states in future states as the consensus goal for agents. On the other hand, RA-CTDE \cite{zhang2024intrinsic} infers the central agent's action tendencies from surrounding agents. However, environmental changes require agents to adjust their consensus continuously, which incurs high computational costs and complexity.

Several communication methods, such as \cite{das2019tarmac,ding2020learning,yuan2022multi,sun2023alice,sun2024assessing,li2025efficient,yao2025general}, design communication networks that enable agents to exchange decision-relevant messages during the decentralized execution.
% design communication networks that enable agents to exchange information during the execution phase.
% For example,
% MAIC \cite{yuan2022multi} optimizes communication among agents by calculating mutual information between trajectories and teammates' actions.
% T2MAC \cite{sun2024t2mac} utilizes evidence theory to enable sparse communication between agents. 
% COCOM \cite{li2025efficient} integrates implicit consensus learning and explicit communication to facilitate collaboration.
However, in some real-world scenarios, limitations such as high noise, high latency, and low bandwidth often prevent these communication algorithms from performing well \cite{9466501,9597491,Song_Lin_Han_Yao_Wu_Wang_Lv_2025}. 
Furthermore, the communication attack may introduce malicious information, disrupting agents' decision-making and hindering collaboration \cite{xue2021mis,zhu2024survey}.
In comparison, our method adopts an alternative perspective. We propose a novel communication-free framework that models teammates' active inference to comprehensively understand their decision logic.
%, thus avoiding the limitations imposed by such environments.

\subsection{Agent Modeling Methods}
% Methods based on agent modeling typically acquire agents' behaviors, beliefs, or intentions \cite{rabinowitz2018machine,yang2018towards,tian2019regularized,ijcai2023p39}. Additionally, \cite{he2016opponent} infers the agent’s policy based on its observations and observed actions. \cite{zintgraf2021deep} models the agent’s policy as a latent distribution.
% % However, relying on incomplete modeling may lead to cognitive bias and negatively impact cooperative performance. 
% However, these methods rely on a strong assumption that agents can directly access the modeled agent' trajectories, which is infeasible during the decentralized execution. 
Methods based on agent modeling typically acquire agents' behaviors, beliefs, or intentions. For example, \cite{rabinowitz2018machine,yang2018towards,tian2019regularized,zintgraf2021deep,ijcai2023p39} model agents’ psychological state and beliefs using the Theory of Mind and Bayesian reasoning. \cite{he2016opponent,raileanu2018modeling} infer agents’ policy based on the modeled agents' observations and actions. \cite{DBLP:journals/corr/abs-2001-10829} model agents’ policy as a latent distribution. 
However, these methods rely on a strong assumption that agents can directly access the modeled agent' trajectories, which is infeasible during the decentralized execution. 
% Methods based on agent modeling typically model incomplete decision processes of other agents, such as policies, beliefs, or intentions \cite{he2016opponent,rabinowitz2018machine,yang2018towards,raileanu2018modeling,tian2019regularized,DBLP:journals/corr/abs-2001-10829,zintgraf2021deep,ijcai2023p39}. However, relying on incomplete modeling may lead to cognitive bias and negatively impact cooperative performance. 
% Meanwhile, these methods rely on a strong assumption that agents can directly access teammates' trajectories, which is infeasible during the execution of CTDE. 

Although \cite{papoudakis2021agent,xie2021learning,yu2024opponent} model other agents solely based on local observations, they maintain teammates' policies as fixed parameters. 
Nevertheless, collaboration performance becomes constrained by fixed teammate parameters, preventing team policy optimization.
% Moreover, the aforementioned methods only partially model other agents' decision-making processes. Such incomplete modeling inevitably introduces inaccuracies due to discrepancies between the model and the actual situation, potentially inducing cognitive biases.
Moreover, the above methods only partially model other agents' decision-making processes, introducing inevitable inaccuracies and negatively impacting cooperative performance owing to mismatches between the model and reality.
In contrast, our framework models teammates' complete decision-making process through three key aspects: perception-belief-action to increase modeling accuracy. In this way, our method enables agents to understand how they think.
% In this way, our method mitigates the uncertainty about teammates in decentralized execution.

\section{Background}

A fully cooperative multi-agent task can be modeled as a \textit{Decentralized Partially Observable Markov Decision Process} (Dec-POMDP) \cite{oliehoek2016concise}, represented as a tuple \(\langle I, S, \{A_i\}_{i=1}^N, \{\Omega_i\}_{i=1}^N, O, \mathcal{T}, R, \gamma \rangle\). 
In this model, \(I = \{1,\dots, N\}\) is the set of agents and \(N\) is the number of agents. 
\(S\) is the state space. 
For each agent \(i \in I\), \(A_i\) is the individual action space, and \(A = A_1 \times \cdots \times A_N\) is the joint action space. 
\(\Omega_i\) is the observation space for agent \(i\). 
\(O(o_i \mid s, i)\) is the observation function over local observations \(o_i \in \Omega_i\) given state \(s \in S\) and agent \(i\). 
% \( \mathcal{O} \) is the observation function
The state transition function \(\mathcal{T}(s' \mid s, \boldsymbol{a})\) defines the probability of transitioning to state \(s' \in S\) given current state \(s \in S\) and joint action \(\boldsymbol{a} \in A\). 
The reward function \(R(s, \boldsymbol{a})\) gives the immediate reward, and \(\gamma \in [0,1)\) is the discount factor, used to balance long-term and short-term rewards. 
At each time step \(t\), agent \(i \in I\) receives a local observation \(o_i \sim O(\cdot \mid s, i)\), selects an action \(a_i \in A_i\), and the joint action \(\boldsymbol{a} = \langle a_1, \dots, a_N \rangle\) results in a state transition and a reward \(R(s, \boldsymbol{a})\). 
The goal of multi-agent system algorithms is to find a joint policy $\boldsymbol{\pi}$  to maximize the expected cumulative reward, formulated as:
$V^{\boldsymbol{\pi}}(s) = \mathbb{E}_{\boldsymbol{\pi}} \left[ \sum_{t=0}^{\infty} \gamma^t R(s_t, \boldsymbol{a}_t) \right]$.
\begin{figure*}[t]
\centering
\includegraphics[width=2\columnwidth]{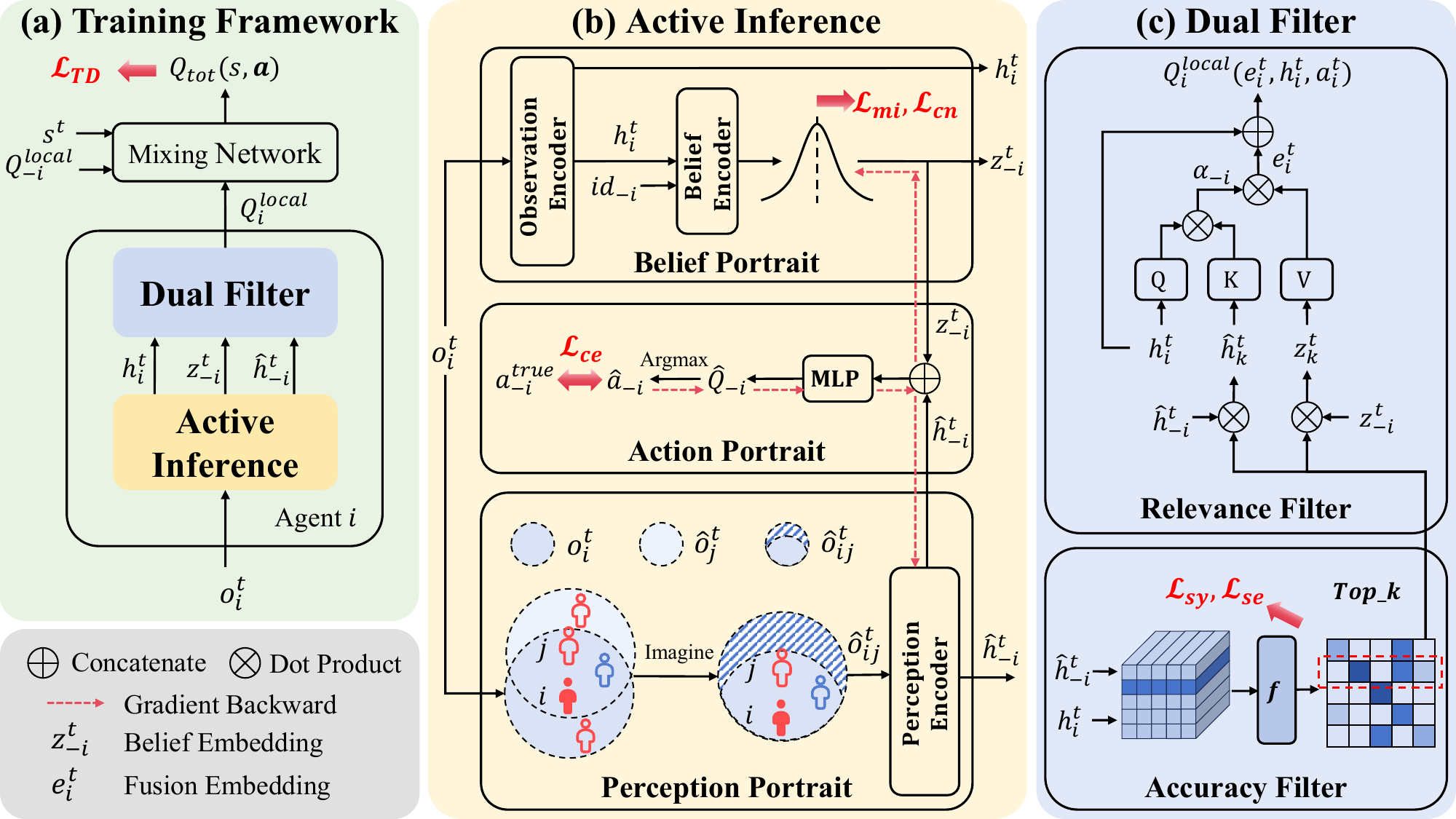}
\caption{The overall framework of AIM. (a) The training framework comprises the agent network and the mixing network; (b) The active inference module, which includes perception portrait, belief portrait, and action portrait; (c) The dual filter module, consisting of the accuracy filter and the relevance filter.}
% AIM的总体框架。(a)训练框架；(b)主动推理模块包括感知建模、信念建模、动作建模；（c）双重过滤模块包括准确性过滤和相关性过滤
\label{fig: framework}
\end{figure*}

\section{Method}

% 我们认为，建模队友的主动推理模型能够提供更多关于队友的信息。然而，由于建模结果可能与真实值存在差异，因此需要评估建模的准确性，并在决策过程中优先使用高准确度的信息。
% PA包含混合网络和智能体网络。核心包括两个方面：一个方面侧重于对队友的意图、观测和的动作进行建模，以弥补部分可观测性造成的信息缺失；一方面基于主体不相关性对队友进行筛选，并基于主体相关性融合队友的建模表示。
% 为了缓解执行阶段全局状态和队友信息缺失带来的决策不确定性，
To build the cognition of teammates' decision logic during decentralized execution, our core idea is to model their complete decision-making process.
% to mitigate information gaps arising from local observations. 
On the one hand, we introduce active inference to construct teammates' three portraits: perception-belief-action, which represent ``how teammates think". On the other hand, due to modeling errors and the diversity of teammates, we devise a dual filter that dynamically integrates teammates' belief portraits based on the accuracy and relevance of their perception portraits.

% PA consists of the mixing network and the agent network. The core includes two aspects: one focuses on modeling teammates' perceptions, beliefs, and actions to compensate for information loss caused by partial observability; the other filters teammates based on agent irrelevance and fuses teammates' modeling representations based on agent relevance.

% \subsection{Dual Modeling Of Teammate Agents}
\subsection{Teammate Portrait via Active Inference}

% 主动推理模型：从感知世界、更新意图、做出决策三个角度讲。

% 我们将建模视角从中心智能体转移到队友，建模队友的主动推理模型，即建模队友观测、信念和动作。在建模过程中，PA根据观测和动作的建模结果，更新队友的建模意图以不断更新队友的主动推理模型。

% 信念代表智能体对环境的更深层次的理解，在一定程度上为智能体的行为趋势提供了依据。

In active inference, the teammate's decision-making process is divided into three stages: perception of the environment, belief formation, and action execution. Perception serves as the foundational step, capturing the change of entities, which supports the following processes. Beliefs represent agents' deeper understanding of the environment, serving as the basis for deriving action. The final decision, which is the action output, is based on both perception and belief and can be used as posterior information to optimize the perceptions and beliefs. Here, we give the details of the triple model process in AIM.
% We shift the modeling perspective from the central agent to the teammate, constructing the teammate's active inference model. This model includes the teammate's perceptions, beliefs, and actions. During the modeling process, PA updates the teammate's beliefs based on the results of perceptions and actions.
% 1、意图建模(为什么要进行意图建模，意图建模的作用是什么)
% 通过对队友的信念进行建模，智能体可以推测队友的目标、意图以及未来可能采取的行动，从而更好地达成合作。
% 为了指导对特定的队友进行建模，对于智能体i，我们使用智能体i的历史轨迹 \( h_i \) 和队友智能体的id-i进行建模，如图a所示,然后将（h_i,id_-i）输入到VAE中，获得队友建模分布N (µti, δti)，通过重参数化，我们得到队友智能体的意图建模表示 \( zi-i \) 。

% 1、观测建模
% PA旨在建模出队友的虚拟局部观测来增强智能体间协作，从而弥补分散执行时全局信息的缺失，减少由于智能体的局部视角造成的不确定性。
% 与 ~\cite{zhang2024intrinsic} 周围智能体预测中心智能体相比，在PA中，对于智能体i，使用中心智能体的oi对队友的观测o-i进行建模。具体过程是通过将每个队友的位置坐标依次设置为原点，重新计算其他智能体相对于该原点的位置坐标，并附加其他相关信息，如图所示。该过程本质上是一个视角转换操作，不依赖任何学习参数。我们只选取oi与计算得到的o-i相交的部分作为建模的结果。然后将o-i作为输入，使用GRU结构获得队友的预测历史轨迹信息\( \hat{h}_{-i}^{t} \)。

% 我们按照图a中对oi的处理对o-i进行相同的操作，得到对队友的想象历史轨迹信息表示h-i，然而，由于o-i只代表了队友的部分观测，所以对于智能体i来说表示h-i是不准确的，我们将对队友建模的表示z-i和h-i拼起来得到想象队友的Q值，并通过最小化想象的Q值和队友真实的Q值Qtrue之间的交叉熵损失来优化想象网络。

% \subsubsection{Observation Modeling}\label{obs model}
% \textbf{Perception Portrait.}
\paragraph{Perception Portrait}
To understand teammates' behavior, agents should first understand what they have experienced. 
AIM aims to construct the world in teammates' eyes, generating their local observations from a teammate-centered perspective.
% AIT aims to model virtual local observations of teammates to enhance cooperation among agents. This compensates for the lack of global information during decentralized execution and reduces the uncertainty caused by the agents' local perspectives.
% Compared with \cite{zhang2024intrinsic}, where surrounding agents predict the central agent, in PA,
In detail, for agent \(i\), the perception \(\hat{o}_{j}^t\) of teammate \(j\) is constructed based on \(i\)'s own observations \(o_i^t\). 
The specific process involves setting each teammate's position as the origin and recalculating the positions of the other agents relative to this origin, while also adding other relevant information.
As shown in Figure \ref{fig: framework}, we only select the portion of \(o_i^t\) that intersects with \(\hat{o}_{j}^t\) as the perception portrait \(\hat{o}_{ij}^t\). 
This process is essentially a viewpoint transformation operation. 
More details about the transformation can be found in  Appendix B. 
We then use $\hat{o}_{ij}^t$ as the input and apply a GRU to obtain the historical trajectory information $\hat{h}_{ij}^{t}$ of the teammate $j$. Subsequently, we acquire $\hat{h}_{-i}^{t}$ of all teammates.

% \subsubsection{Intent Modeling}
\paragraph{Belief Portrait}
% \textbf{Belief Portrait.}
% By modeling a teammate's beliefs, an agent can infer the teammate's goals and intentions, thereby facilitating better collaboration. 
% Belief serves as a higher-level abstraction of actions, reflecting the agent's behavioral trends. 
% Belief characterizes agents' posterior distributions over hidden states, serving as the core basis for optimal policies.
Belief serves as a higher-level abstraction of actions, serving as the core basis for policies. 
However, unlike the perception portrait, which is objective, the belief portrait typically exhibits high subjectivity and variability.
In scenarios with limited observation, these characteristics become more pronounced, making it challenging to model beliefs from teammates' perspectives accurately. 
% Additionally, the subsequent decision-making is agent-centric, and the belief portrait needs to be interpretable to the agent.

Therefore, we construct the belief portrait of teammates from the agent's perspective. 
For agent \( i \), we use its trajectory \( h_i^{t} \) and teammates' index \(id_{-i}\) for modeling, as shown in Figure \ref{fig: framework}. The input \( (h_i^{t}, id_{-i}) \) is fed into the belief encoder to obtain the belief distribution \( \mathcal{N}(\mu_{i}^{t}, \delta_{i}^{t}) \). Through reparameterization, we obtain the teammates' belief portraits \( z_{-i}^{t} \) which should exhibit two characteristics: 
(1) Decision-support ability;
(2) Stability over the short term.

% 在队友的主动推理过程中，队友的动作由信念产生。为了得到对队友的强大建模表示z，我们采用在MAIC中提出的互信息loss来优化模型，mi_loss是队友采取的动作与以hi和id-i为条件的z之间的互信息。

% 在队友的主动推理过程中，我们认为，给定队友的历史信息 \( h_-i \) 和 \( i_{d-i} \)的情况下，队友的动作是由信念 \( z \) 产生的。 \( z \) 与队友采取的动作之间存在一定的依赖关系。在此基础上，为了得到对队友的强大建模表示 \( z \)，我们引入了互信息损失（mi_loss），该损失度量了队友的动作与信念 \( z \)（条件于 \( h_i \) 和 \( i_{d-i} \)）之间的互信息。

% 通过条件于队友的历史信息 \( h_i \) 和独立信息 \( i_{d-i} \)，我们推断出队友的信念 \( z \)。队友的行为 \( a_j \) 为我们提供了优化信念的关键信息。为了使信念表示 \( z \) 更加符合队友的真实信念，我们引入互信息损失 \( \text{mi\_loss} \)，它度量了队友的动作与信念表示 \( z \) 之间的先验依赖关系。通过最大化互信息损失，我们能够优化信念 \( z \)，使其更准确地反映队友的决策过程。
To enhance the decision-support ability, we get inspired by \cite{yuan2022multi} and maximize the mutual information between teammates' actions and belief portraits \( z_{-i}^{t} \), conditioned on \( h_i^{t} \) and \(id_{-i}\). Through variational inference, maximizing mutual information can be transformed into the following loss.
The derivation can be found in Appendix A.
\begin{equation}\label{mi_loss}
\begin{aligned}
\mathcal{L}_{mi} = & \, \mathbb{E} \left[\mathcal{D}_{KL}\left( p(z_{-i}^{t} \mid h_i^{t}, id_{-i}) \parallel \right. \right. \\
& \, \left. \left. q_\xi(z_{-i}^{t} \mid h_i^{t}, a_{-i}^{t}, id_{-i}) \right) \right],
\end{aligned}
\end{equation}
% 其中分布p和q的变量是从从经验池buffer中采样得到，DKL表示KL散度，最大化mi_loss相当于最小化以智能体的局部消息为条件的队友模型的不确定性.
where \( \mathcal{D}_{KL}(\dots||\dots) \) represents the Kullback-Leibler divergence, \( q_\xi(z_{-i}^{t} \mid h_i^{t}, a_{-i}^{t}, id_{-i}) \) is used as a variational distribution to approximate the conditional distribution \( p(z_{-i}^{t} \mid h_i^{t}, id_{-i}) \). 
Minimizing $\mathcal{L}_{mi}$ is equivalent to maximizing the relevance between the belief portraits and 
selected actions.

% 由于 \( z \) 是对队友智能体的意图建模表示，预计其在短时间内会保持一定的连续性，因此，我们通过计算相邻时间步队友表示 \( z \) 之间的余弦相似度，并通过最大化这一相似度来提升队友表示的稳定性和一致性,公式化为：
To improve the stability of \( z_{-i}^{t} \), we calculate the cosine similarity between the \( z_{-i}^{t} \) at adjacent time step, formalized as:
\begin{equation}\label{continuity_loss}
\mathcal{L}_{cn} = \mathbb{E} \left[ - \frac{z_{-i}^{t-1} \cdot z_{-i}^t}{\| z_{-i}^{t-1} \| \| z_{-i}^t \|} \right].
\end{equation}

% 意图反映了智能体的长期规划，而队友的动作揭示其即时反应；通过动作建模，智能体能够在当前状态下更好地理解队友的策略，并基于队友的即时行为信息快速调整自己。
% 对于队友的主动推理模型中的信念更新过程，我们对队友的动作进行建模来更新信念建模的表示。

% 在我们的方法中，使用Q值进行动作队友的信念。建模，将z-i和h-i拼起来得到队友的想象动作Q值，并最小化想象动作Q值和队友的真实动作Q值Qtrue之间的交叉熵损失来优化动作建模模块。此外，联合动作会造成环境状态转移从而影响队友的感知画像。
% \subsubsection{Action modeling}
\paragraph{Action Portrait}
% \textbf{Action Portrait.}
% Through action modeling, the agent can better understand the teammate's policy in the current state and quickly adjust its own behavior based on the teammate's real-time behavior information.
% In the belief updating process of the teammate's active inference model, we model the teammate's actions to update the representation of the belief.
Action is the most direct outcome of active inference. The accuracy of the action serves as a critique of the modeling process, especially the coherence of the belief portrait. 
% Moreover, the state transitions resulting from joint actions can influence the perception portrait of teammates. 
Moreover, joint actions induce environmental state transitions, which in turn influence teammates' perception portraits.
Hence, we optimize the perception and belief portrait by leveraging posterior action predictions.
In AIM, we concatenate belief portraits \( z_{-i}^{t} \) and historical perception information \( \hat{h}_{-i}^{t} \) as the input of action prediction network, and obtain the imagined action distribution \( \hat{a}_{-i} \). By minimizing the cross-entropy loss between \( \hat{a}_{-i}\) and the true actions \(a_{-i}^{true} \), AIM optimizes the action modeling network and the perception-belief portrait.
\begin{equation}\label{ce_loss}
\mathcal{L}_{ce} = - \sum_{i} a_{-i}^{true} \log \hat{a}_{-i}.
\end{equation}

% 队友智能体双重建模部分的损失表示为：
% \begin{equation}
% \mathcal{L}_{model} = \lambda_{mi} \mathcal{L}_{mi} + \lambda_{cn} \mathcal{L}_{cn} + \lambda_{ce} \mathcal{L}_{ce}.
% \end{equation}
% 其中，\lambda_{mi},\lambda_{cn}和\lambda_{ce}分别表示互信息损失，连续性损失和交叉熵损失的超参数，来平衡它们的效果。
The complete loss for the teammate portrait via active inference is expressed as:
\begin{equation}\label{model_loss}
\mathcal{L}_{MD} = \lambda_{mi} \mathcal{L}_{mi} + \lambda_{cn} \mathcal{L}_{cn} + \lambda_{ce} \mathcal{L}_{ce},
\end{equation}
where \( \lambda_{mi} \), \( \lambda_{cn} \), and \( \lambda_{ce} \) represent the hyperparameters of the loss function, which are used to balance their effects.

%合作队友选择
\subsection{Dual Filter of Accuracy and Relevance}
% \subsection{Selection And Fusion}
% 由于队友建模结果与真实值之间存在差异，筛选出建模准确度较高的队友对于实现有效合作至关重要。同时，考虑到建模的是队友的主动推理过程，如何利用建模过程中提取的信息来优化自身决策，也是一个值得思考的问题。在这一部分，我们将重点解决这两个核心问题。

Due to local observation, errors in the active inference process are unavoidable. Indiscriminately utilizing erroneous portraits of teammates can directly distort the agent's comprehension of the current environment, resulting in non-cooperative behavior. Furthermore, since collaboration in multi-agent systems is often localized, it is redundant to incorporate the portraits of all teammates in decision-making. 
Hence, we propose a dual filter mechanism for selecting cooperative teammates, focusing on two aspects: the accuracy and the relevance of portraits.
% In the triple portrait, only the perception portrait is objective and error-free, with differences arising solely in terms of completeness.
% Due to the discrepancies between modeling portraits and real teammates, it is crucial to select teammates with higher modeling accuracy for effective collaboration. 
% At the same time, considering that we model the teammate's active inference process, how to leverage the information extracted during this modeling process is also a question worth deeper exploration.

% In this section, we select the top $k$ teammates to cooperate with based on the teammate observation modeling \( \hat{h}_{-i}^{t} \), and make decisions using their action representations \( z_{k}^{t} \).

% %合作队友初步筛选
% 在局部可观测条件下使用非交互方法而不是用自注意力机制选择通信对象的原因和优点。
% 拟真性
% 如果预测的观测与真实的接近，则矩阵值高，说明预测的观测与真实的观测接近，对智能体建模的动作表示越真，隐式消息越可靠。

% 正如我们之前提到的，对队友的观测建模是视角变化的结果，对于队友来说，建模的观测内容是独立的，与其他智能体无关，不涉及智能体间交互。同时由于观测建模结果与队友真实的观测之间存在差异，因此我们考虑筛选出拟真性较高的队友来促进协作。
% 为了评估智能体 \( i \) 对每个队友建模的准确度，我们将观测建模得到的历史轨迹 \( h_{-i} \) 输入到双层 MLP，并通过 softmax 操作得到对每一个队友建模的拟真性矩阵 \( C \)。

% \subsubsection{Agent-irrelevant Realism fliter}
% 观测建模是视角变化的结果。对于每一个主体智能体，队友的观测建模结果是以队友自身为中心得出的。
% 这些建模结果本质上是智能体自身观测的一个子集，且与其他智能体无关。同时由于只是子集，观测建模的结果是片面的。因此需要一个评价网络来评价观测建模结果的拟真性，同时该网络应该具有以下性质：
% 1、“自己是保真的”。自身智能体建模自身观测的结果肯定是真实的，所以网络得出的评价应该是最高的；
% 2、“评价是相似的”。对于智能体i，智能体i对队友j观测建模的评价应该与队友j对智能体i的评价相似。
% 3、“高相似性赋予高分”。如果一个智能体的观测建模结果和其他任何智能体的真实观测相似，说明该结果具有高拟真性，那么网络应给予该结果较高的评价。
% % 为了实现这一目的，我们通过将h-i和真实的hi输入两层MLP并通过softmax操作构建一个n×n的评价矩阵C，其中n表示智能体数量。
\paragraph{Accuracy Filter}
Among triple portraits, the perception portrait is derived through perspective transformation, rendering it inherently partial.
% For each agent, the observation modeling results of teammates are obtained with the teammates themselves as the center. 
% These results are essentially a subset of the agent's observations and are irrelevant to other agents.
Therefore, an evaluation method is required to assess the accuracy of perception portraits.

Specifically, we learn a mapping $f: \mathbb{R}^h \mapsto \mathbb{R}$ that maps the perception portrait to an accuracy score. At time $t$, we simultaneously process the $N\times N$ portraits, constructing the evaluation matrix $\mathcal{C}^t$. $N$ represents the number of agents. $c^t_{ij}$ refers to the accuracy score of agent $i$'s perception portrait $\hat{h}_{ij}^{t}$ to agent $j$, formed as:
% In order to fulfill this, we input the observation representations \( \hat{h}_{-i}^{t} \) and the real observations \( h_i^t \) into a two-layer MLP. Then, we apply the softmax operation to construct an \( \mathcal{N} \times \mathcal{N} \) evaluation matrix $\mathcal{C}$, where \( \mathcal{N} \) represents the number of agents. 
\begin{equation}\label{softmax}
c^t_{ij} = \text{softmax}(f(\hat{h}_{ij}^{t})),
\end{equation}
% 其中fj(.)表示两层MLP网络，矩阵值cij表示智能体i对智能体j建模的拟真性分数，值越高表示智能体i对智能体j建模的越真实。我们将根据拟真性矩阵C选出k个建模最真实的队友作为合作对象。
where \( f(\cdot) \) represents a two-layer MLP network. A higher value of \( c_{ij} \) signifies that the perception portrait of agent $j$ by agent $i$ is more accurate. The evaluation matrix $\mathcal{C}^t$ should possess the following characteristics.

(1)\textbf{Mutual evaluation is similar.} For agents \( i \) and \( j \), perception portraits of each other are essentially different perspectives on the same intersection of their observations.
% For agent \( i \), agent \( i \)’s evaluation of teammate \( j \)’s perception modeling should be similar to teammate \( j \)’s evaluation of agent \( i \).

(2)\textbf{Self-evaluation is highest.} The perception portrait of an agent's own is guaranteed to be accurate, so the network's evaluation of this result should be the highest.

(3)\textbf{High similarity gets a high score.} If an agent’s perception portrait is similar to the real observation of any other agent, it indicates high accuracy. 

% 为了满足我们提到的评价的对称性，我们引入对称性约束来优化评价矩阵，公式化为
To satisfy the \textbf{Mutual evaluation is similar}, we introduce a symmetry loss $\mathcal{L}_{sy}$ to optimize the evaluation matrix $\mathcal{C}$, which is formalized as:
% We introduce two matrix constraints to optimize the teammate selection module. First, we consider that cooperation is mutual. Specifically, from the perspective of agent $i$, if agent $i$ needs to cooperate with agent $j$, then from agent $j$'s perspective, agent $j$ should also consider the influence of agent $i$. Therefore, we compute the symmetric value of the realism matrix for modeling as $\mathcal{L}_{c}$ to constrain this process.
\begin{equation}\label{symmetry_loss}
\mathcal{L}_{sy} = \| \mathcal{C} - \mathcal{C}^T \|_\mathcal{F},
\end{equation}
% 其中CT是矩阵C的转置
where $\mathcal{C}^T$ is the transpose of the matrix $\mathcal{C}$, and $\| \cdot \|_\mathcal{F}$ denotes the Frobenius-norm. 

% 另外，为了满足自己是保真的，我们引入对角线损失来优化评价网络。
To enhance \textbf{Self-evaluation is highest}, we utilize a diagonal loss $\mathcal{L}_{se}$ to maximize the accuracy score of true $h_{ii}^t$.
% Furthermore, for agent $i$ to cooperate with a teammate, it is essential that agent $i$ clearly understands its own objectives. That is, agent $i$ should not focus solely on the policy of the teammate and forget its own task goals. Therefore, we compute the trace of the matrix $\mathcal{C}$ as $\mathcal{L}_{self}$, which constrains the realism matrix.
\begin{equation}\label{self_loss}
\mathcal{L}_{se} = -\sum_{i} c_{ii}.
\end{equation}

% 对于\textit{“High similarity gets a high score”}，我们设计的双层MLP能够通过其非线性转换捕捉输入之间的相似性。由于其结构特点，相似的输入会通过网络的激活函数产生较高的输出，从而满足该性质。
Regarding \textbf{High similarity gets a high score}, deep neural networks exhibit the property that similar inputs produce similar outputs. Combined with characteristic (2), perception portraits similar to the true $h_{ii}^t$ can be mapped to higher scores, thereby satisfying characteristic (3) without the additional loss function.
% This is because of continuous and smooth activation functions and non-linear transformations between layers, thereby satisfying this property.
% the designed two-layer MLP can capture the similarity between inputs through its nonlinear transformation. Due to its structural characteristics, similar inputs are processed by the network's activation functions to produce higher outputs, 

% 然而考虑与所有的队友达成合作往往会带来信息的冗余，在评价矩阵C中按照拟真性的值优先级采样k个索引作为合作对象的索引。
Having the evaluation matrix $\mathcal{C}$, we sample the \(top\_k\) indices for each agent based on the accuracy to select potential teammates for the next filter.

% Then, drawing from the concept of experience replay \cite{schaul2015prioritized}, we sample \( k \) indices as potential cooperation partners based on their realism values in the matrix, prioritizing those with higher realism.

% % 隐式消息融合（相关性）
% 在这一部分，为了使智能体能够将目光进一步集中，我们采用注意力机制对筛选后的k个队友智能体表示信息进行再次提取，注意力机制因为其可以通过动态地分配不同信息源的重要性，而广泛应用在多源信息融合中。在PA中，我们将智能体真实的hi作为询问，选取的k个队友智能体的观测建模表示hk作为key，k个队友智能体的动作建模表示zk作为value，对k个合作队友的建模表示进行融合，注意力分数表示为：

% \subsubsection{Fusion of {\textquotedblleft implicit message\textquotedblright}}
% \subsubsection{Agent-relevant Realism Fusion}
% 队友的意图的作用，根据感知层的相似性对意图进行融合，解释使用意图而不是用动作的原因
% 我们直观上认为意图表示的是智能体的长期规划，理解队友的意图往往比了解队友的单步具体动作更为重要。通过理解队友的意图，智能体能够预判队友未来的行为，从而维持长期有效的合作。另外，历史信息相似表明智能体和队友可能有相似的经验，这使得智能体能够根据队友的信念模型进行有效的融合。因此，在这部分，我们考虑根据观测建模结果对队友的意图进行融合。
% 相比于了解队友单步的动作，了解队友的信念可以使智能体了解队友的长期规划，从而预判队友未来的行为，维持长期有效的合作。
% \begin{figure*}[ht]
% \centering
% \includegraphics[width=1\columnwidth]{figure/overview_results_smac_1.pdf}
% \caption{Performance comparison between AIM and baselines on nine representative maps in SMAC.}
% % AIT和baselines在SMACv2和GRF中的性能对比,第一行是GRF的结果，第二行是SMACv2的结果
% \label{fig: smac}
% \end{figure*}

\begin{figure*}[ht]
	\centering
	\includegraphics[width=1\linewidth]{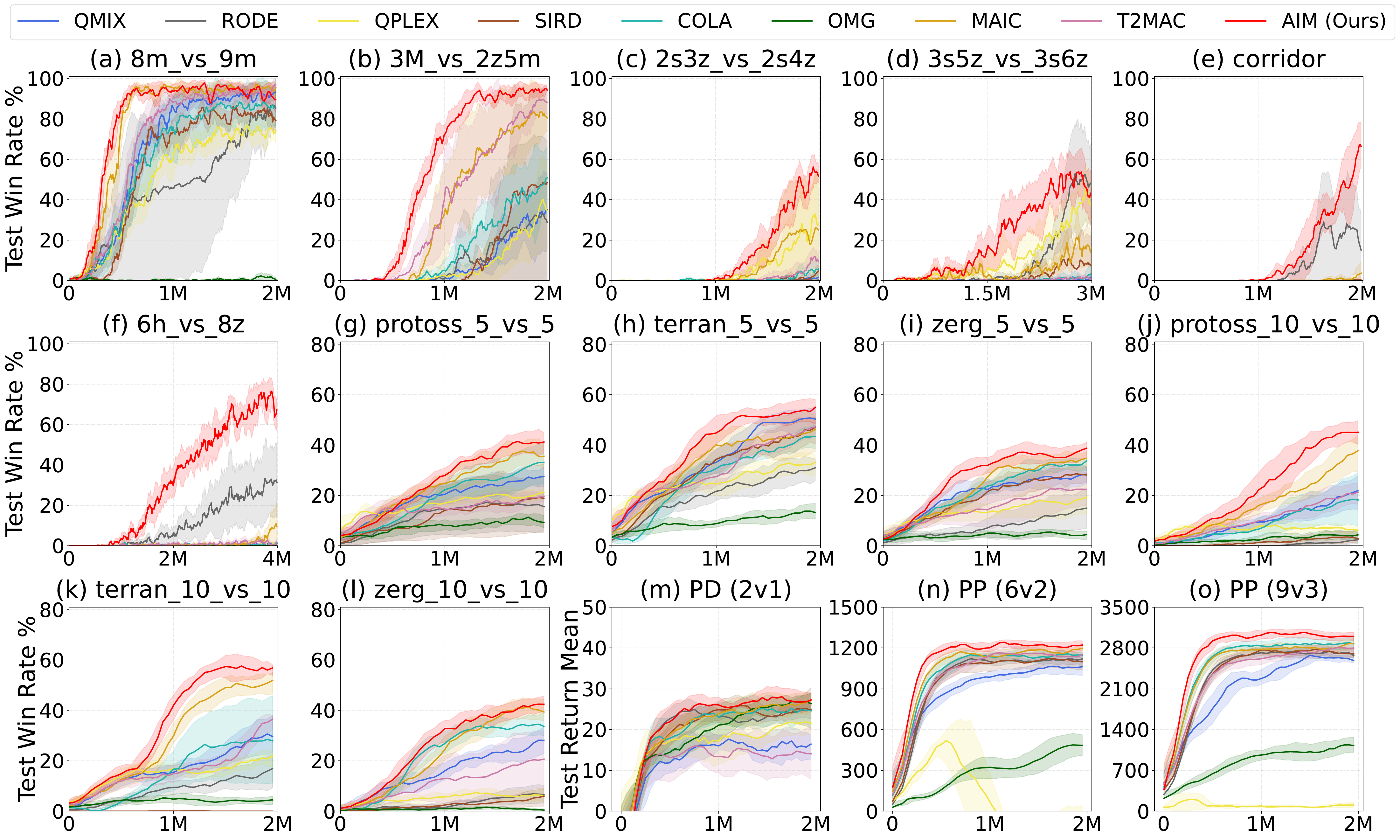}
	\caption{Performance comparison between AIM and baselines on SMAC, SMACv2, and MPE. (a)-(f) Six representative maps on SMAC. (g)-(l) Six tasks on SMACv2. (m)-(o) Three tasks on MPE.}
	\label{fig: smac}
\end{figure*}

% 另外，观测的相似性反映的是轨迹历史信息的相似性，考虑与自己历史信息相似的队友的信念可以预测队友未来的潜在行动。这使得智能体能够提前调整决策，从而更好地协调与队友的合作。
\paragraph{Relevance Filter}
% We intuitively consider that intentions represent an agent's long-term plan, and understanding a teammate's intentions is often more important than understanding their immediate actions. By understanding a teammate's intentions, an agent can anticipate their future behavior, thereby maintaining long-term and effective collaboration. Furthermore, knowing the intentions of teammates whose perceptions are similar to own can reduce the uncertainty in own policy. Therefore, in this section, we consider integrating teammates' intentions based on the observation modeling results.
Building on the accuracy filter, we also need to apply the relevance filter from a decision-making perspective, selecting teammates more relevant to the agent for effective collaboration.
Given that collaboration in multi-agent systems is often localized, we use the perception portrait as a proxy for relevance.
As for how to utilize teammates' portraits to aid decision-making, the most intuitive approach is to combine teammates' action portraits to mitigate the lack of cognition about teammates. However, due to local observation, accurately modeling actions is challenging. 
% Using incorrect teammate actions to guide one's own decisions is often inefficient. 
Therefore, we instead combine belief portraits, leveraging higher-level behavioral bases to dilute the impact of single-step modeling errors.
% Compared to understanding a teammate's immediate observations or actions, understanding their beliefs enables an agent to comprehend a teammate's long-term plan. Moreover, the similarity of observations reflects the similarity of trajectory history. By considering the beliefs of teammates whose historical information is similar to one's own, an agent can predict their potential future actions. This enables the agent to adjust its decisions in advance. Therefore, in this section, we consider integrating teammates' beliefs based on the perception modeling results.
% 为了实现这一目标，PA采用注意力机制对意图建模表示进行融合。注意力机制因为其可以通过动态地分配不同信息源的重要性，而广泛应用在多源信息融合。在PA中，我们将智能体真实的hi作为询问，选取的k个队友智能体的观测建模表示hk作为key，k个队友智能体的意图建模表示zk作为value，对k个合作队友的zk进行融合，注意力分数表示为：

Here, we apply the attention mechanism \cite{vaswani2017attention} to achieve the relevance filter and fusion of belief portraits. 
% The attention mechanism is widely used in multi-source information fusion because it dynamically assigns different levels of importance to various information sources. 
In AIM, for agent \( i \), we adopt the true perception history \( h_i^{t} \) as the query, the historical perception \( \hat{h}_k^{t} \) of the selected $k$ teammates as the key, and adopt the belief portrait representations \( z_k^{t} \) of the $k$ teammates as the value. We then fuse the \( z_k^{t} \), with the attention score represented as:
\begin{equation}\label{att_score}
\alpha_{i,k} = \frac{\exp \left( \frac{1}{\sqrt{d_{key}}} \, (h_i^{t} W_Q) \cdot (\hat{h}_k^{t} W_K)^T \right)}{\sum_{j=1}^{k} \exp \left( \frac{1}{\sqrt{d_{key}}} \, (h_i^{t} W_Q) \cdot (\hat{h}_j^{t} W_K)^T \right)},
\end{equation}
where \( W_Q \) and \( W_K \) are learnable weight matrices applied to the query \( h_i^{t} \) and the key \( \hat{h}_k^{t} \), while \( d_{key} \) denotes the dimension of the key vector.
% 注意力分数表示智能体i和合作队友智能体j之间的相关性，为了最大限度的减少智能体i决策的不确定性，我们从MAIC中引入一个熵正则化项，并最小化该项：
The attention score $\alpha_{i,k}$ represents the relevance between agent \(i\) and its collaborative teammates. 
% To enhance the focus of attention, we introduce an entropy regularization term from \cite{yuan2022multi} and aim to minimize it.
% \begin{equation}\label{lr_loss}
% \mathcal{L}_{er} = \sum_{i=1}^n \mathcal{H}(\alpha_i \cdot) = - \sum_{i \neq k}^n \alpha_{ik} \log \alpha_{ik}.
% \end{equation}
% 最终的注意力结果为ci,并将其与hi一起馈送到线性网络得到局部Q值Qi-local
The combined result is \(e_i^{t} = \sum_{j=1}^{k} \alpha_{i,j} \cdot z_j^{t}\), which is concatenated with \( h_i^{t} \) and subsequently processed by a linear network to compute the local \(Q\)-value \( Q_i^{local} \).

% 双重队友选择部分的优化目标为：
The optimization objective of this section is:
\begin{equation}\label{l_se}
\mathcal{L}_{DF} = \lambda_{sy} \mathcal{L}_{sy} + \lambda_{se} \mathcal{L}_{se},
\end{equation}
% 其中，\lambda_{cp},\lambda_{self},\lambda_{r}表示损失的超参数
where \( \lambda_{sy} \) and \( \lambda_{se} \) represent the hyperparameter of the loss function.

% 总体优化目标：
\subsection{Overall Training Objective}
% 队友建模表示融合后，我们采用最基础的CTDE训练架构QMIX进行训练，同时也可以扩展到任何像MAPPO和MADDPG的CTDE框架。所有的模型参数通过全局Q值Qtot的TD损失进行更新，同时PA还可以扩展到其他的值分解方法，如VDN，QPLEX等：
% 完整的训练损失如下：
Following the fusion of teammates' belief portraits, we utilize the basic CTDE training framework QMIX \cite{rashid2020monotonic} for training. 
% The pseudocode of our method can be found in Appendix C.3.
Meanwhile, AIM remains compatible with other value decomposition methods, such as VDN \cite{sunehag2017value} or QPLEX \cite{wang2020qplex}.
% Meanwhile, our method can be extended to other CTDE frameworks, such as MAPPO \cite{yu2022surprising} or MADDPG \cite{lowe2017multi}. 
All model parameters are updated by minimizing the $\mathcal{L}_{TD}$:
\begin{equation}
\mathcal{L}_{TD} = \mathbb{E} \left[ \left( y - Q_{tot}(\bm{\tau}, \bm{\mathit{a}}) \right)^2 \right],
\end{equation}
where $y=r + \gamma \max_{a'} \hat{Q}_{tot}(\bm{\tau'}, \bm{\mathit{a'}})$ is the target network of the joint action-value function.
% \tau \sim \mathcal{D}} \left[ \left( r + \gamma \max_{a'} \hat{Q}_{tot}(s', \bm{\mathit{a'}})
By integrating the triple portrait and the dual filter, the complete training loss is as follows:
\begin{equation}\label{tot_loss}
\mathcal{L}_{tot} =\mathcal{L}_{TD} + \mathcal{L}_{MD} + \mathcal{L}_{DF}.
\end{equation}

\section{Experiments}

\begin{figure*}[t]
\centering
\includegraphics[width=1\linewidth]{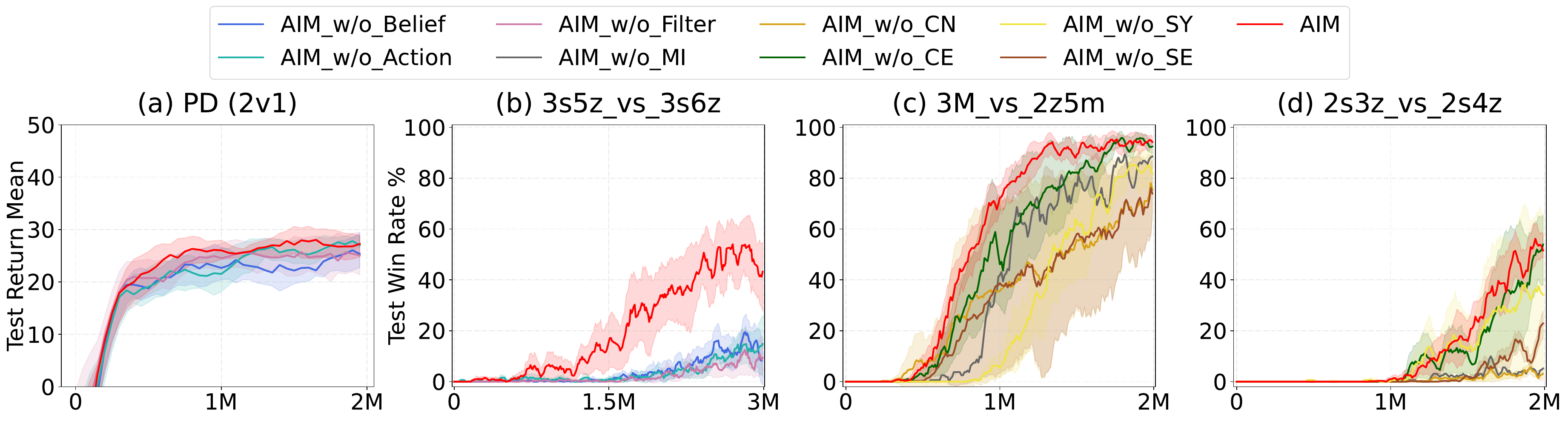}
\caption{Ablation studies. (a)-(b) illustrate module-wise ablation. (c)-(d) present loss-wise ablation. ``AIM\_w/o\_Belief" removes belief portrait, ``AIM\_w/o\_Action" removes action portrait, while ``AIM\_w/o\_Filter" removes dual filter.}
% AIM和baselines在SMACv2和GRF中的性能对比,其中，sight=0.8表示智能体视野半径为0.8，sight=1.2表示智能体视野半径为1.2。(a)-(b)表示逐个模块消融。(c)-(d)表示逐loss消融。(e)-(h)表示对不同参数k的对比。
\label{fig: ablation_loss}
\end{figure*}

% 我们选择了基于CTDE（Centralized Training with Decentralized Execution）的方法QMIX、QPLEX、COLA、ROMA、RODE和SIRD作为我们的主要基准算法（baseline）。同时，我们将OMG中的友方智能体设置为均需要训练的以作为智能体建模方法基准。此外，为了评估AIM建模框架与流行的通信算法之间的差异，我们还选择了两个通信算法MAIC和T2MAC作为额外的基准。值得注意的是，AIM是一个非通信框架，因此，将其与大量通信算法进行性能比较是不可取的。同时，我们选用OMG作为智能体建模baseline。我们在实验中主要回答以下几个问题：（1）PA的效果能否能超过baseline方法（2）AIM 提出的主动推理模块能否帮助缓解执行阶段队友信息的缺失？（3）PA中的筛选队友过程能否过滤掉冗余信息？（4）PA能否通过拟真性矩阵筛选出合作队友？
% 为了评估AIM在缓解执行阶段队友不确定性的能力，我们还选择了两个通信算法MAIC和T2MAC作为额外的基准。
% We select six CTDE-based methods, QMIX \cite{rashid2020monotonic}, QPLEX \cite{wang2020qplex}, COLA \cite{xu2023consensus}, ROMA \cite{wang2020roma}, RODE \cite{wang2021rode}, and SIRD \cite{zeng2023effective}, along with two communication algorithms, MAIC \cite{yuan2022multi} and T2MAC \cite{sun2024t2mac}, as our baselines. 
We select several methods as our primary baselines, including QMIX \cite{rashid2020monotonic}, QPLEX \cite{wang2020qplex}, RODE \cite{wang2021rode}, COLA \cite{xu2023consensus}, and SIRD \cite{zeng2023effective}. 
% To evaluate the performance between AIM and communication-based methods, we include MAIC \cite{yuan2022multi} and T2MAC \cite{sun2024t2mac} as supplementary baselines. 
To evaluate the efficacy of our communication-free method in constructing teammates' decision logic, we include two communication-based methods: MAIC \cite{yuan2022multi} and T2MAC \cite{sun2024t2mac} as supplementary baselines. 
% It is worth noting that AIM is a non-communication framework, so direct performance comparisons with numerous communication-based methods are inappropriate. 
Furthermore, we set OMG \cite{yu2024opponent} with all teammates being trainable, serving as a baseline for agent modeling.

\subsection{Performance Comparisons}
% \subsubsection{SMAC}
% For question \textbf{(1)}, we conduct experiments on three benchmarks: SMAC \cite{samvelyan2019starcraft}, SMACv2 \cite{ellis2024smacv2}, and Google Research Football (GRF) \cite{kurach2020google}. To eliminate the impact of dead agents on the teammate selection process, we mask the dead agents during the training of cooperative teammate selection in both SMAC and SMACv2.

% For question \textbf{(1)}, 
We conduct experiments on four MARL benchmarks: SMAC \cite{samvelyan2019starcraft}, SMACv2 \cite{ellis2024smacv2}, the Multi-agent Particle-Environment (MPE) \cite{mordatch2018emergence} and the Google Research Football (GRF) \cite{kurach2020google}.
% To mitigate the impact of dead agents on the accuracy and relevance filter of teammates, we apply a mask to dead agents during the training phase in both SMAC and SMACV2.
% 值得注意的是，虽然GRF是一个全观测的环境，但是踢足球的任务需要智能体之间进行有效的协作才能完成进球得分，对智能体建模方法的协作性提出了挑战。我们主要在该环境中评估AIM达成的协作效果。
% To evaluate AIM's universality in fully observable environments and its non-stationarity mitigation capacity, we incorporate the Google Research Football (GRF) \cite{kurach2020google} as an additional benchmark. Notably,  despite GRF guarantees global state observations, persistent non-stationarity emerges from agents' dynamic policy interactions.
Notably, although GRF provides a fully observable environment, the soccer task demands frequent collaboration to achieve scoring goals, posing challenges to the agent modeling methods. 
% We primarily evaluate the collaborative performance of AIM in this setting.
% 同时，考虑到AIM的通用性，我们希望AIM也能够适用于全观测的环境，因此我们选用更加具有挑战的Google Research Football (GRF)环境来评估AIM。
For evaluation, we use five different random seeds and plot the average test results as bold lines. 
% Throughout the experiments, we ensure consistent environmental settings across all methods. 
% Detailed configurations of the environment hyperparameters are provided in Appendix C. 
Due to page limitations, we leave the GRF experiments in Appendix D.1.

\paragraph{SMAC}
% SMAC is a popular benchmark in MARL, where agents need to learn various micro-operations, such as "focus fire" or "kite", in order to defeat enemies. As shown in Figure. \ref{fig: smac}, our method, AIM, performs well across many maps, especially on super-hard maps such as 3M\_vs\_2z5m, 2s3z\_vs\_2s4z, and 6h\_vs\_8z. In complex scenarios where success depends on clear task division and collaboration, selecting teammates who can effectively coordinate is useful for improving the chances of success.
As the widely used benchmark in MARL, SMAC requires agents to master micro-level policies such as ``focus fire," ``kite," and ``draw aggro" to defeat enemies under relatively concentrated initial positions. 
% To evaluate the performance of AIM, we select eight representative maps that cover three difficulty levels: easy, hard, and super hard. 
% Super hard maps often feature complex tasks that require agents to finely tune their policies for effective cooperation.
% Additionally, in environments with a large number of agents, each agent must identify the best collaboration partners to maintain effective cooperation. 
% Since AIM is designed to model the thought processes of teammates and incorporates selective cooperation, it is essential to validate its performance and advantages on these maps.
% Figure. \ref{fig: smac} presents the results of AIM across eight tasks. AIM achieves optimal or near-optimal performance on most maps.
Figure \ref{fig: smac} (a-f) presents the performance of AIM on six tasks, demonstrating optimal results on most maps.
Particularly on maps that require a clear division of labor and the selection of the best collaborators, such as \texttt{3s5z\_vs\_3s6z}, \texttt{corridor}, and \texttt{6h\_vs\_8z}, AIM even outperforms communication-based baselines. 
This demonstrates that when agents are nearby, communication exerts negligible effects on decision-making quality. In contrast, AIM enables agents to select the most cooperative teammates by modeling teammates' active inference processes. This capability enhances task allocation efficiency and contributes to achieving team objectives. In our OMG setup, each allied agent independently models teammates, failing to capture inter-agent coordination patterns and thus impairing collaboration.
% All results of AIM in SMAC are presented in Appendix C.2.
% , and Appendix D.2 shows the experimental results and settings of MAPPO-based AIM.
% 基于MAPPO的AIM实验结果和实验设置可以在附录中找到

% 我们设置的OMG中每个友方智能体都是独立对队友进行建模，无法有效获得智能体间协作模式，削弱了协作效果。

% SMACv2
% 作为SMAC的扩展，SMACV2引入了初始位置和单位类型的随机性，同时智能体位置较分散，进一步提高了对算法的要求。在我们的实验中，我们设置了10个盟友与10个敌人对抗，AIM与基准算法的性能如图2所示。由于难以适应环境的随机性，基准表现较差。相比之下，AIM能够更好地适应这种不确定性环境。实验结果表明，AIM中提出的观测画像模块能够有效适应智能体的观测范围变化，验证了AIM在建模观测画像方面的有效性。同时，即使SMACv2的智能体初始位置分散，AIM也能达到甚至高于基于通信的baseline。

% \subsubsection{SMACv2} 
\paragraph{SMACv2}
% SMACv2 builds upon SMAC by introducing randomness in the starting positions, unit types, unit vision, and attack range, which raises the requirements for the algorithm. In our experiment, we set up 10 allies against 10 enemies. Figure. \ref{fig: GRF_smacv2} illustrates the performance of AIM across three maps in SMACv2. In environments filled with randomness, integrating teammates' beliefs allows the agent to understand their long-term plans, rather than focusing on their short-term actions or observations.
As an extension of SMAC, SMACv2 introduces additional randomness in the initial positions and unit types, with dispersed agent distributions. It also employs true unit attack and sight ranges.
% that intensify algorithmic demands.
% in starting positions, unit types, unit observation ranges, and attack ranges
% In our experiments, we set up a scenario where 10 allies face off against 10 enemies, and 
The performance of AIM is shown in Figure \ref{fig: smac} (g-l). Due to the difficulty in adapting to the environmental randomness, the baseline methods perform poorly. In contrast, AIM demonstrates superior adaptability to this uncertain environment. The results show that the proposed perception portrait module in AIM effectively adjusts to changes in the agents' sight ranges, validating its efficacy. Furthermore, even with initially dispersed agent positions, AIM achieves performance comparable to or surpassing communication-based baselines. This validates the efficacy of AIM in assisting agents to understand teammates' decisions and enhancing cooperation.

% 在我们的实验中，我们选用

% GRF
% 我们进一步评估了 AIM 在具有挑战性的 GRF 环境中的表现。GRF 是一项足球基准测试，其特点是部分可观察性更加严峻，需要频繁、高效的传球才能实现有效的团队协调和得分。由于GRF具有高度稀疏的奖励机制，只有在进球时才会给予+1的奖励，训练过程较为困难。为了加速训练，我们采用了与COLA相同的环境设置，并确保所有基准算法使用与我们相同的环境配置。实验结果如图所示，AIM在该环境中表现优于所有基准算法。这表明，在具有强部分可观测性且频繁动作交互的复杂环境中，AIM能够通过建模队友的主动推理过程，准确预测其未来行为模式，从而显著提升团队的协作与协调能力。
% \subsubsection{GRF} 

% \begin{figure*}[ht]
% 	\centering
% 	\includegraphics[width=1\textwidth]{figure/overview_results_smacv2.pdf}
% 	\caption{Performance comparison between AIM and baselines on six representative maps on SMACv2.}
% 	\label{fig: smacv2}
%     % AIM在3M_vs_2s5z环境中训练的策略生成的测试轨迹场景可视化。下方展示了对应时间步的感知准确性矩阵，其中，亮度越高表示感知画像的准确性越高。方框标示了被选中的队友索引。
% \end{figure*}

% \begin{figure}[t]
% \centering
% \includegraphics[width=1\columnwidth]{figure/overview_results_belief.pdf}
% \caption{Analysis of belief portraits from different perspectives. ``AIM (Teammate)'' models belief portraits from the teammate's perspective, while AIM adopts the agent's perspective for modeling.}
% % 信念画像不同视角分析建模分析.``AIM(Teammate)"表示从队友的视角构建使用队友的感知画像对信念画像进行建模.而我们的方法AIM从智能体视角对队友的信念画像进行建模.
% \label{fig: belief}
% \end{figure}

\paragraph{MPE}
The MPE provides a 2D physics-based environment supporting both continuous and discrete action spaces \cite{papoudakis2020benchmarking}.
We evaluate the performance of AIM in the following three tasks with discrete actions: (1) Physical-Deception(PD)(2v1), (2) Predator-Prey(PP)(6v2), and (3) Predator-Prey(PP)(9v3).
% \texttt{(1)Physical-Deception(2v1)}
% % : Two agents cooperatively identify and occupy the authentic target landmark through coordinated coverage. These agents need to actively deceive a heuristic-controlled agent attempting to detect and approach the target. 
% \texttt{(2) Predator-Prey(6v2)}
% % : Six predators coordinate to capture two heuristic-controlled agents. 
% \texttt{(3) Predator-Prey(9v3)}
% % : Nine predators coordinate to capture three heuristic-controlled agents.
% Detailed environment specifications are provided in Appendix C.2.
% MPE是一个基于物理的2D环境，其中动作空间可以是连续的也可以是离散的（epymarl）。我们选用MPE中的三个任务，包括Physical deception、Predator-Prey和Cooperative navigation来评估AIM在离散动作空间的MPE中的性能。在Physical deception中我们控制两个智能体尽可能的覆盖目标地标并迷惑一个试图识别接近目标地标的启发式智能体。在Physical deception中我们控制拥有三个Predator的团队来抓捕一个内置的启发式智能体，在Cooperative navigation中我们控制三个智能体在避免碰撞的前提下导航至三个地标。详细的环境介绍可以在附录中找到。
% 考虑到原始的MPE环境并不是部分可观测的，为了aim感知画像的有效性，我们在范围为[-1,1]的环境中将智能体的视野范围设置为0.8和1.2，分别对应hard和easy难度。图4的实验结果表明，即使是在强部分可观测下，aim也可以通过建模队友的主动推理过程了解队友的决策逻辑，从而促进协作。由于页数限制，我们将Cooperative navigation的实验结果放在附录中。
Since the original MPE is fully observable, we modify it to evaluate the effectiveness of the perception portrait. Specifically, we set the agents' view radius to 0.8. 
% Detailed environment specifications are provided in Appendix C.2.
% The results are shown in Figure \ref{fig: GRF_smacv2}. Compared to baseline methods, AIM facilitates coordination by modeling the teammates' active inference process and understanding their decision-making.
As demonstrated in Figure \ref{fig: smac} (m-o), AIM effectively infers teammates' decision logic through modeling, enabling collaboration even under strict partial observability conditions.

\begin{figure*}[t]
\centering
\includegraphics[width=1\linewidth]{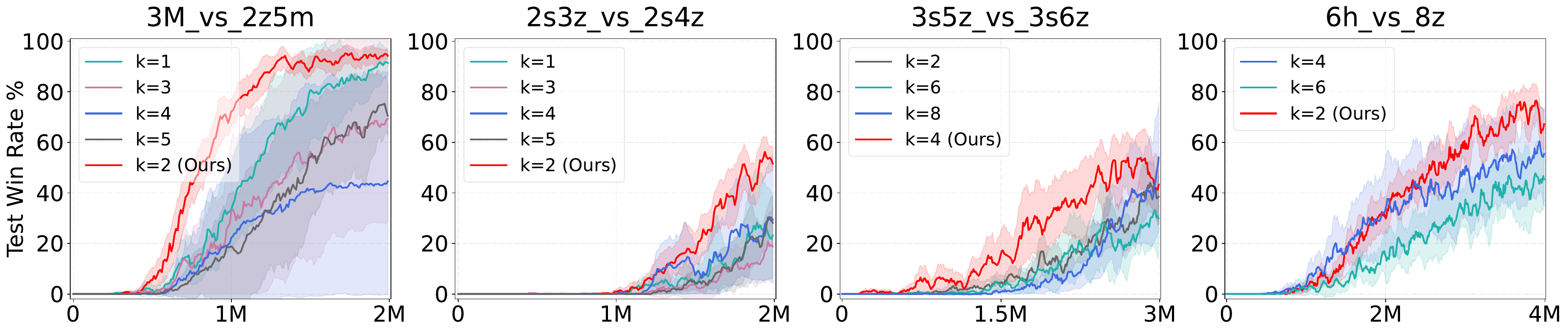}
\caption{Analysis of the parameter \( k \) in selective collaboration. Different values of \( k \) have varying impacts on performance.}
% AIM和baselines在SMACv2和GRF中的性能对比,其中，sight=0.8表示智能体视野半径为0.8，sight=1.2表示智能体视野半径为1.2。(a)-(b)表示逐个模块消融。(c)-(d)表示逐loss消融。(e)-(h)表示对不同参数k的对比。
\label{fig: ablation_k}
\end{figure*}

\begin{figure*}[t]
\centering
\includegraphics[width=1\linewidth]{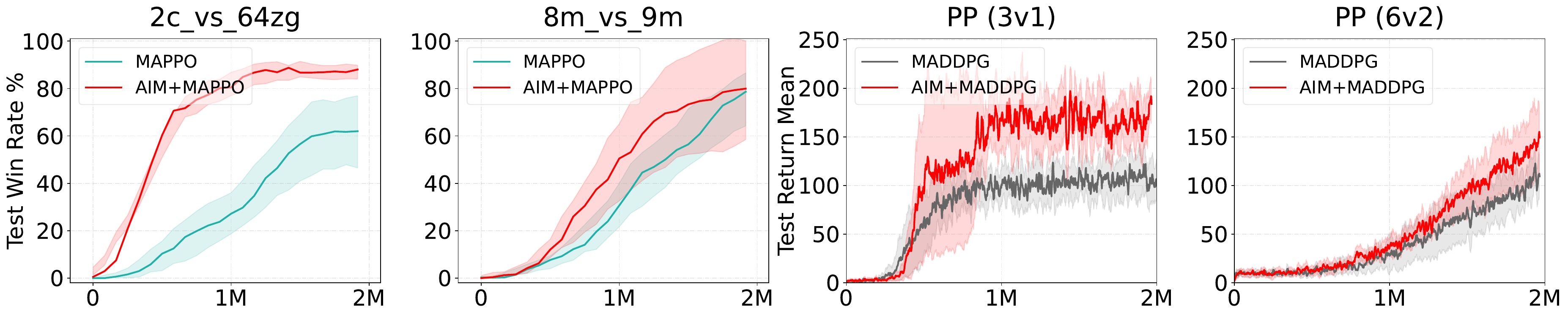}
\caption{Scalability Experiments. AIM + MAPPO denotes the extension of AIM to the MAPPO framework, while AIM + MADDPG represents the adaptation of AIM to the MADDPG framework for continuous action spaces.}
% 可扩展性实验。其中，AIM+MAPPO表示将AIM扩展至MAPPO框架，AIM+MADDPG表示将AIM扩展至适用于连续环境的MADDPG框架。
\label{fig: scalability}
\end{figure*}

% 消融实验
\subsection{Ablation Studies}
% 在这一部分，我们通过消融实验深入分析AIM中的主动推理模块和双重过滤机制对整体性能的影响。具体来说，我们通过逐步移除这些关键模块，观察其对系统性能的影响，以评估它们在复杂任务中的作用。
% In this section, we primarily analyze the impact of AIM's active inference module and dual-fliter mechanism on overall performance through ablation experiments.
In this section, we analyze the impact of active inference and dual filter modules on overall performance using ablation experiments. We progressively remove these key modules and evaluate their effects.  
% \textbf{This allows us to assess each module's role in handling complex tasks.}
% 为了验证AIM主动推理模块在解决由于队友信息缺失带来的决策不确定性问题中的有效性，我们进行了消融实验。由于SMACV2的随机观测范围已验证了AIM感知画像模块的效果，我们重点分析信念画像模块和动作画像模块。我们将AIM与四种消融方式进行了对比：1）AIM_w/o_Belief，仅消除信念画像模块中的LCN和LSE；2）AIM_w/o_Action，仅消除动作画像模块中的LCE；3）AIM_w/o_ALL,消除信念画像和动作画像；4）QMIX。图1中的消融结果清晰表明，无论消去哪个模块，都会导致AIM性能的下降。而在完全删除这两个模块后，性能显著下降，但仍优于QMIX。表明这两个模块对AIM的整体性能均起到了重要作用。此外，单独移除信念画像模块时，性能下降更为明显。表明在复杂任务中，了解队友的信念对于有效协作至关重要。另外，图4中的(c)-(d)表明移除每一个损失函数都会造成性能下降，表明每一个损失对AIM
% 此外，我们进一步分析了从客观角度对队友信念画像建模的有效性，我们选择了更具挑战性的地图 3s5z_vs_3s6z。如图3所示，结果表明，从智能体视角构建的队友信念画像能够有效减少不确定性。而如果使用队友主观且不准确的感知画像对信念画像进行建模，则会导致一定的认知偏差，进而影响协作效果。为了分析AIM中的损失函数的影响，我们对每个损失函数进行了消融。
% 为了验证从智能体角度构建信念画像的有效性，我们选择了更具挑战性的地图 \texttt{3s5z\_vs\_3s6z}。如图 \ref{fig: belief} 所示，从代理的角度构建队友信念画像可以有效降低不确定性。相反，依赖从队友角度构建的不准确的感知画像会引入阻碍协作的认知偏见。
% 如图 \ref{fig: belief} 所示，AIM (Teammate)依赖从队友角度构建的不准确的感知画像来建模信念画像会引入阻碍协作的认知偏见。相反，AIM从代理的角度构建队友信念画像可以有效降低不确定性。
% For \textbf{(2)}, we remove the \( \mathcal{L}_{cn} \) and \( \mathcal{L}_{mi} \) from belief modeling, and the \( \mathcal{L}_{ce} \) from action modeling, with the results shown in Figure. \ref{fig: ablation}. When either of these modules was removed, the algorithm performance decreased, indicating that modeling teammates' beliefs are crucial for understanding their behavior patterns. Additionally, continuously adjusting belief modeling and perception modeling based on action modeling proves to be essential.

\paragraph{Active Inference Module}
We conduct ablation studies to validate the active inference module's efficacy.
% To validate the effectiveness of the active inference module in mitigating decision-making uncertainties arising from missing teammate information during execution, we conduct the ablation study. 
% Since the different observation range of SMACv2 has already confirmed the efficacy of the perception portrait
Since the different sight ranges among units of SMACv2 have already validated the efficacy of the perception portrait, we focus on analyzing the belief and action portrait. 
We compare AIM with three ablation configurations: 1) \texttt{AIM\_w/o\_Belief}, which only removes the belief portrait; 2) \texttt{AIM\_w/o\_Action}, which only removes the action portrait; 3) \texttt{AIM\_w/o\_Filter}, which removes the dual filter of accuracy and relevance. 
% The ablation results in Figure \ref{fig: ablation_loss} (a)-(b) demonstrate that removing any module leads to a decline in performance. This highlights the critical role both portraits and dual filters play in enhancing overall performance.
The ablation results presented in Figure \ref{fig: ablation_loss} (a)-(b) demonstrate that the removal of any individual module leads to significant performance degradation. This highlights that: (1) the comprehensive modeling of teammates' decision processes is essential for maintaining accurate models, and (2) the dual filters are crucial for obtaining optimal teammates' information.

To analyze the impact of different loss functions in AIM, we conduct ablation studies on each loss function. As demonstrated in Figures \ref{fig: ablation_loss} (c)-(d), the elimination of any single loss function results in significant performance deterioration. 
% These results show that each loss plays an indispensable role in maintaining the overall effectiveness. 
Furthermore, the performance decline is more pronounced when $\mathcal{L}_{cn}$ or $\mathcal{L}_{se}$ is removed alone. 
This highlights the critical importance of maintaining continuous belief modeling of teammates while preserving self-focused decision cognition for enhancing complex task success rates.
% This highlights the essential need for sustained teammates' belief modeling while maintaining self-focused cognition during complex task execution.
% This underscores the critical importance of understanding teammates' beliefs while maintaining self-focus for effective collaboration in complex tasks.
% Figures \ref{fig: ablation_loss} (c)-(d) further demonstrate that removing any individual loss function from AIM leads to significant performance degradation, validating the essential role of each loss component in AIM. 

% Figures \ref{fig: ablation_loss} (c)-(d) further demonstrate that removing any individual loss function from the AIM framework leads to significant performance degradation, validating the essential role of each loss component in AIM. 
% Furthermore, the performance decline is more pronounced when the part of the belief portrait is removed alone, underscoring the importance of understanding teammates' beliefs for effective collaboration in complex tasks.
% When both belief and action portraits are completely removed, performance drops significantly but remains superior to QMIX.
% This highlights the critical role both portraits play in enhancing overall performance. Furthermore, the performance decline is more pronounced when the belief portrait is removed alone, underscoring the importance of understanding teammates' beliefs for effective collaboration in complex tasks.

\begin{table}[t]
    \centering
    \begin{tabular}{cccc}
        \hline
        Maps  & QMIX & AIM(Teammate)  & AIM(Ours)\\
        \hline
        8m\_vs\_9m     & 87.2     & 89.1     & \textbf{94.0} \\
        3M\_vs\_2z5m     & 32.3     & 84.4     & \textbf{93.3}  \\
        3s5z\_vs\_3s6z  & 00.0     & 50.4     & \textbf{58.7} \\
        6h\_vs\_8z  & 00.0     & 53.6     & \textbf{67.6} \\
        \hline
    \end{tabular}
    \caption{Analysis of belief portraits from different perspectives. ``AIM (Teammate)'' models belief portraits from the teammates' perspective, while AIM adopts the agent's perspective for modeling. Bold indicates the best performance.}
    \label{tab:belief}
\end{table}

Subsequently, we compare the construction of belief portraits from the agent's perspective with that from the teammate's perspective.
% As shown in Figure. \ref{fig: belief}, the results demonstrate that constructing teammate belief portraits from the agent's perspective effectively reduces uncertainty. In contrast, using subjective and inaccurate perception portraits from teammates to model belief portraits introduces cognitive biases, ultimately hindering collaboration performance.
Table \ref{tab:belief} shows the \texttt{AIM (Teammate)} models belief portraits based on inaccurate perception portraits constructed from teammates' perspectives. This approach introduces modeling biases that hinder collaboration. In contrast, AIM constructs belief portraits of teammates from the agent's perspective, effectively reducing inaccuracies and improving cooperative performance.

% As shown in Figure. \ref{fig: belief}, adopting the agent's perspective to model teammates' belief portraits effectively reduces uncertainty. In contrast, relying on inaccurate perception portraits based on the teammate's perspective introduces cognitive biases that hinder collaboration.
% 对于AIM中的双重过滤模块，我们分析了从感知准确性矩阵中选取的参数 \(k\) 的影响，重点探讨了双重过滤在冗余信息筛选中的作用。我们选取了SMAC中的两个超难度地图进行实验，实验结果如图2所示。可以观察到，参数 \(k\) 的选择对AIM的性能有显著影响。随着 \(k\) 的增大，智能体能够与更多的队友达成合作，协作带来的信息量增大，从而显著提升了任务完成的成功率。然而，当 \(k\) 超过某一阈值时，一些队友的信息开始产生冗余，且不准确的消息进入智能体的决策过程，导致性能下降。因此，双重过滤模块可以帮助智能体合理地选择合作队友的数量，从而有效避免信息冗余和决策中的误导，优化任务表现。为了帮助理解双重过滤中准确性过滤的作用，我们在3M_vs_2z5m中对准确性过滤中的评价矩阵进行了可视化，并放在了附录D中。

% \begin{figure}[t]
% \centering
% \includegraphics[width=1\columnwidth]{figure/overview_results_k.pdf}
% \caption{Analysis of the parameter \( k \) in selective collaboration. Different values of \( k \) have varying impacts on performance.}
% % 选择性协作中参数k的分析。不同的k对效果有不同的影响。
% \label{fig: k}
% \end{figure}

% To answer \textbf{(3)}, we further analyze the parameter $k$. As shown in Figure. \ref{fig: k}, the experimental results demonstrate that the choice of $k$ affects the model's performance. As $k$ increases, the number of teammates cooperating with the agent also increases, leading to a noticeable improvement in performance. However, when \( k \) exceeds a certain threshold, redundant and inaccurate information begins to flood the agent's decision-making process, leading to degraded performance.
\paragraph{Dual Filter Module}
We analyze the impact of the parameter \(k\) selected from the perception evaluation matrix, with a focus on its role in filtering redundant information. We conduct experiments using four super-hard maps from SMAC, and the results are shown in Figure \ref{fig: ablation_k}.
It is evident that the choice of \(k\) significantly influences the performance of AIM.
As \(k\) increases, collaboration with more teammates enhances task success rates due to increased information, but surpassing a certain threshold introduces redundancy, hindering decision-making. 
% This results in a decline in performance.
% Therefore, the dual filter module assists the agent in selecting the optimal number of collaboration partners, effectively preventing information redundancy and misleading decisions.
Therefore, the dual filter module optimizes the teammate selection, effectively preventing information redundancy and misleading decisions.
% To clarify the effectiveness of the dual filtering module, we analyze the effect of the accuracy matrix within the dual filter module by visualizing the \texttt{3M\_vs\_2z5m} task as shown in Appendix E.
% To facilitate comprehension of the accuracy filter module, we visualize the evaluation matrix in the \texttt{3M\_vs\_2z5m} scenario and present it in Appendix E.

\subsection{Scalability Across Policy-Based Frameworks}
% 为了进一步研究AIM在其他CTDE框架下的可扩展性，我们将AIM扩展至MAPPO和适用于连续动作的MADDPG。详细的实验和参数设置见附录D。我们在SMAC和具有连续动作的MPE环境中进行了评估。结果如图3所示，AIM+MAPPO的效果要优于MAPPO，AIM+MADDPG的效果要优于MADDPG。这表明AIM在其他CTDE框架中具有良好的可扩展性，同时，其在具有连续动作空间的环境中也展现出优异的扩展能力。
To further investigate the scalability of AIM across different policy-based frameworks, we extend AIM to both MAPPO \cite{yu2022surprising} and MADDPG \cite{lowe2017multi}. 
% Detailed experimental configurations and parameter settings are provided in Appendix D. 
We conduct evaluations in both SMAC and continuous-action MPE. The results in Figure \ref{fig: scalability} demonstrate AIM + MAPPO surpassing MAPPO and AIM + MADDPG exceeding MADDPG. 
% The performance gain stems from triple portrait modeling and dual filtering, mitigating information deficits.
% The performance gain stems from information compensation via triple portraits and partner selection via dual filters.
% The performance gain stems from the triple portrait modeling that compensates for teammates' information, coupled with selective collaboration enabled by the dual filters.
Experimental results validate the dual scalability capabilities of AIM: (1) extensibility within policy-based frameworks, and (2) adaptability to continuous action space environments.

\section{Conclusion and Discussion}

% 在这篇文章中，我们聚焦于多智能体协作问题，仅考虑对队友进行建模,提出了一种名为AIM的队友建模框架，以缓解CTDE执行阶段的非平稳性。该框架通过建模队友主动推理过程的感知画像、信念画像和动作画像来了解队友的思考方式，并通过双重过滤模块剔除准确性和相关性较低的画像。实验结果与消融实验充分验证了AIM的优越性。然而，在需要考虑选择性协作的场景中，AIM通过选择固定的top_k个队友作为合作伙伴，如何根据画像动态选择协作队友仍然是一个值得进一步研究的问题，我们将其留待未来工作探索。
% In this paper, we focus on modeling teammates and propose AIM, a framework that reduces uncertainty about teammates during the execution of CTDE.
% In this paper, we replace ``\textit{Tell}" with ``\textit{Think}" to obtain teammates' information and propose AIM, a framework that reduces uncertainty about teammates during the execution of CTDE.
In this paper, we propose AIM, a novel framework that replaces explicit communication (i.e., ``\textit{Tell}'') with agent modeling (i.e., ``\textit{Think}'') to construct the cognition of teammates' decision logic.
% In this paper, we consider exclusively the modeling of teammates. We propose a teammate modeling framework named AIM to address non-stationarity during the execution phase of CTDE. 
% AIM models teammates' active inference by three portraits: perception, belief, and action, providing deeper insights into teammates' decision-making process. 
% Additionally, AIM introduces the dual filter module to remove portraits with low accuracy and relevance.
% Experiments and ablation studies demonstrate the superiority of AIM. 
AIM models teammates' active inference through perception, belief, and action portraits, and employs a dual filter module to exclude inaccurate and irrelevant portraits. Experiments validate the effectiveness of AIM.
However, in scenarios where selective collaboration is required, AIM selects a fixed set of \(top\_k\) teammates as collaborators. The dynamic selection of collaborative teammates based on portraits remains an open issue and is left for future work.

\section{Acknowledgments}
This work was supported by the National Natural Science Foundation of China (Grant No. 62576029) and the Aeronautical Science Foundation of China (Grant No. 202300010M5001).
% AAAI is especially grateful to Peter Patel Schneider for his work in implementing the original aaai.sty file, liberally using the ideas of other style hackers, including Barbara Beeton. We also acknowledge with thanks the work of George Ferguson for his guide to using the style and BibTeX files --- which has been incorporated into this document --- and Hans Guesgen, who provided several timely modifications, as well as the many others who have, from time to time, sent in suggestions on improvements to the AAAI style. We are especially grateful to Francisco Cruz, Marc Pujol-Gonzalez, and Mico Loretan for the improvements to the Bib\TeX{} and \LaTeX{} files made in 2020.

% The preparation of the \LaTeX{} and Bib\TeX{} files that implement these instructions was supported by Schlumberger Palo Alto Research, AT\&T Bell Laboratories, Morgan Kaufmann Publishers, The Live Oak Press, LLC, and AAAI Press. Bibliography style changes were added by Sunil Issar. \verb+\+pubnote was added by J. Scott Penberthy. George Ferguson added support for printing the AAAI copyright slug. Additional changes to aaai2026.sty and aaai2026.bst have been made by Francisco Cruz, Marc Pujol-Gonzalez, and Mico Loretan.

% \bigskip
% \noindent Thank you for reading these instructions carefully. We look forward to receiving your electronic files!

\bibliography{aaai2026}

\end{document}